\documentclass[useAMS,usenatbib,usegraphicx]{mn2e}

\usepackage{paper_def}
\usepackage{txfonts}
\usepackage[T1]{fontenc}
\usepackage{aecompl}
\usepackage{color}
\title[ALMA observations of normal galaxies at z$\sim$7]{The assembly of ``normal'' galaxies at
z$\sim$7 probed by ALMA}
\author[Maiolino et al.]{R.~Maiolino,$^{1,2}$
S.~Carniani,$^{1,2,3}$ A. Fontana,$^{4}$
L.~Vallini,$^{5,6}$ L.~Pentericci,$^4$ A.~Ferrara,$^5$
\newauthor
E.~Vanzella,$^7$
A.~Grazian,$^4$ S.~Gallerani,$^5$ M.~Castellano,$^4$ S. Cristiani,$^8$
G.~Brammer,$^9$
\newauthor
 P.~Santini,$^4$ J.~Wagg$^{10}$ and R.~Williams$^{1,2}$
\\
$^{1}$Cavendish Laboratory, University of Cambridge, 19 J. J. Thomson Ave., Cambridge CB3 0HE, UK\\
$^{2}$Kavli Institute for Cosmology, University of Cambridge, Madingley Road, Cambridge CB3 0HA, UK\\
$^{3}$Dipartimento di Fisica e Astronomia, Università di Firenze, Via G. Sansone 1, I-50019, Sesto Fiorentino
(Firenze), Italy\\
$^{4}$INAF - Osservatorio Astronomico di Roma, via Frascati 33, 00040 Monteporzio, Italy\\
$^{5}$Scuola Normale Superiore, Piazza dei Cavalieri 7, 56126 Pisa, Italy\\
$^{6}$Dipartimento di Fisica e Astronomia, Università di Bologna, viale Berti Pichat 6/2, 40127 Bologna, Italy\\
$^{7}$INAFâBologna Astronomical Observatory, via Ranzani 1, I-40127 Bologna, Italy\\
$^{8}$INAF- Trieste Astronomical Observatory, via Tiepolo 11, 34143 Trieste, Italy\\
$^{9}$Space Telescope Science Institute, 3700 San Martin Drive, Baltimore, MD  21218, U.S.A.\\
$^{10}$Square Kilometre Array Organization, Jodrell Bank Observatory, Lower Withington, Macclesfield, Cheshire
SK11 9DL, UK\\
}
\begin{document}

\date{Accepted . Received}

\pagerange{\pageref{firstpage}--\pageref{lastpage}} \pubyear{2002}

\maketitle

\label{firstpage}

\begin{abstract}
We report new deep ALMA observations aimed at investigating the
[CII]158$\mu$m line and continuum emission in three spectroscopically confirmed
Lyman Break Galaxies
at 6.8$<$z$\le$7.1, i.e. well within the re-ionization epoch.
With Star Formation Rates of $\rm SFR\sim 5-15~M_{\odot}~yr^{-1}$ these systems are
much more representative of the high-z galaxy population
than other systems targeted in the past by millimeter observations.
For the galaxy with the deepest observation we detect [CII] emission at redshift z=7.107,
fully consistent with the Ly$\alpha$ redshift, but
spatially offset by 0.7$''$ (4~kpc) from the optical emission.
At the location of the optical emission, tracing both the Ly$\alpha$ line and the far-UV continuum,
no [CII] emission is detected in any of the three galaxies,
with 3$\sigma$ upper limits significantly lower than the [CII] emission
observed in lower reshift galaxies.
These results suggest 
that molecular clouds in the central parts of primordial galaxies are rapidly
disrupted by stellar feedback. As a result, 
[CII] emission mostly arises from more external accreting/satellite clumps of neutral gas.
These findings are in agreement with recent models of galaxy formation.
Thermal far-infrared continuum is not detected in any of the three galaxies. However,
the upper limits on the infrared-to-UV emission
ratio do not exceed those derived in metal-- and dust--poor galaxies.
\end{abstract}


\begin{table*}
 \centering
% \begin{minipage}{140mm}
  \caption{Sample and parameters of resulting from the observations.}
  \begin{tabular}{lccccccccccc}
  \hline
   Name     & RA(J2000) & DEC(J200) &  z$_{\rm Ly\alpha}$         & SFR$^a$ & $\rm \nu _{obs}([CII])^b$ &
   beam$^c$ & t$_{on}$(int.)$^d$ & antennae & $\sigma _{\rm cont}^e$ &  $\sigma _{line}^f$ & L([CII])$^g$ \\
     & [deg] & [deg]  &            & [$\rm M_{\odot}~yr^{-1}$] & [GHz] & [min$''\times$maj$''$
	   (PA$^{\circ}$)] & [hours] & & [$\rm \mu$Jy] & [mJy] & [$\rm 10^7~L_{\odot}$]\\
 \hline
 BDF-3299 & 337.0511 & $-$35.1665 & 7.109$^h$ & 5.7 & 234.374 & $\rm 0.55 \times 0.76~(85)$ & 5.1 & 25--36 & 7.8 & 0.062 & $<$2.0 \\
 BDF-512  & 336.9444 & $-$35.1188 & 7.008$^h$ & 6.0 & 237.330 & $\rm 0.72 \times 0.50 ~(68)$ &  1.4 & 29 & 17.4 & 0.171 & $<$6.0 \\
 SDF-46975 & 200.9292 & $+$27.3414 & 6.844$^i$ &  15.4 & 242.290 & $\rm 0.96 \times 0.82~(79)$ & 2.0 & 40 & 19.2 & 0.173 & $<$5.7 \\
 \hline
\end{tabular}
\\
Notes:$^a$This is the SFR inferred from the rest-frame UV continuum and
is adapted to the new calibration given in \citet{Kennicutt2012}, who adopt a Kroupa IMF.
$^b$Expected redshifted frequency of [CII] (LSRK) from the Ly$\alpha$ redshift.
$^c$This column gives the synthesized beam (median value across SW1)
minor and major axes (in arcsec) and position angle (in degrees, anticlockwise
relative tot the North).
$^d$On-source integration time (in hours).
$^e$Rms on the continuum.
$^f$Rms measured in the spectrum extracted from the central beam, avoiding regions of higher noise associated with
atmospheric absorption (except for SDF-46975, see text), in spectral channels of 100~km/s.
$^g$The upper limits on the [CII] luminosities are at 3$\sigma$, and are calculated on a width of 100~km/s. 
$^h$From \cite{Vanzella2011}.
$^i$From \cite{Ono2012}.
%\end{minipage}
\label{tab_list_sources}
\end{table*}

\section{Introduction}

Millimeter and submillimeter observations of distant galaxies are a powerful tool
to trace the evolution of the gas and dust content \citep[e.g.][]{Tacconi2013,Santini2014,Bothwell2013,Genzel2010a,
Daddi2010b,Carilli2013a},
to investigate their dynamics \citep[e.g.][]{Wang2013,De-Breuck2014,Carniani2013,Carilli2013b,Willott2013,Willott2015b}
to trace obscured star formation \citep[e.g.][]{Swinbank2014,Gallerani2012,Negrello2010,Hezaveh2013,Schaerer2015b}, and even to trace
AGNs and their effect onto their host galaxy \citep{Maiolino2012,Cicone2015,Gallerani2014}.

Molecular (mostly CO) transitions are used to trace the content of molecular gas, while continuum
emission is used to trace warm dust heated by ongoing star formation (and possibly by an AGN).
At z$>$4 mm/submm
observations are even more effective since the luminous [CII]158$\mu$m 
far-infrared fine structure line is redshifted into the primary bands of atmospheric transmission, at
frequencies $\rm \nu < 350 GHz$
\citep{Maiolino2005,Maiolino2009,Walter2009,De-Breuck2011}. The [CII]158$\mu$m line is one of the main coolants of the ISM, both
in local and high-z galaxies, generally excited by heating of UV photons in star forming regions,
primarily associated with their Photo-Dissociation-Regions \citep{Madden1997,Kaufman1999,Gracia-Carpio2011,Pineda2014},
although contribution from diffuse atomic and partially ionized gas, as
well as from shocked gas, has been
inferred in some galaxies \citep{Pineda2014,Appleton2013,Velusamy2014,Cormier2012}.

Until recently, at z$>$4 [CII], CO and continuum emission had
been detected only in extreme galaxies such as quasar hosts or SMGs, characterized by star formation
rates of the order of 1000~$\rm M_{\odot}~yr^{-1}$, certainly not representative of the bulk of the
galaxy population. In the last few years a few galaxies at z$\sim$4--6
with more modest star formation rates
($\rm 20-300~M_{\odot}~yr^{-1}$) have been detected in [CII] or in continuum 
\citep{Carilli2013b,Wagg2012,Carniani2013,Williams2014,Riechers2014}.
It should be noted that even these SFR are not yet really representative of the bulk of galaxyies
in the early Universe, which have SFR typically lower than 10 $\rm M_{\odot}~yr^{-1}$ \citep{Salvaterra2011,Finkelstein2012,Dayal2014,Robertson2015}.
At z$\sim$7, well
within the re-ionization epoch, only one quasar host galaxy has been detected in [CII] and in
continuum \citep[at z=7.08, ][]{Venemans2012}. ``Normal'' galaxies at z$\sim$7,
with star formation rates more representative of
the galaxy population at these early epochs, have not yet been detected either in
[CII] or in continuum \citep{Ota2014,Schaerer2015a,Gonzalez-Lopez2014}. We also note that z$\sim$7.7 is
the maximum redshift at which galaxies are spectroscopically confirmed \citep{Oesch2015} and we recall that, even
with the broad ALMA band, a spectroscopic redshift is needed to observe far-IR transitions redshifted
into the millimeter band.

We have undertaken an ALMA programme targeting a sample of three spectroscopically confirmed galaxies
at $\rm 6.8<z<7.1$ with the aim of detecting or constraining both the [CII] line
and the continuum thermal emission.

For consistency with other papers on similar sources we assume the following cosmological parameters:
$\rm H_0 = 70~km~s^{-1}~Mpc^{-1}$, $\rm \Omega_\Lambda = 0.7$
and $\rm \Omega _m = 0.3$.

\begin{figure*}
\centerline{\includegraphics[width=6.5truecm,angle=90,bb=150 53 430 729]{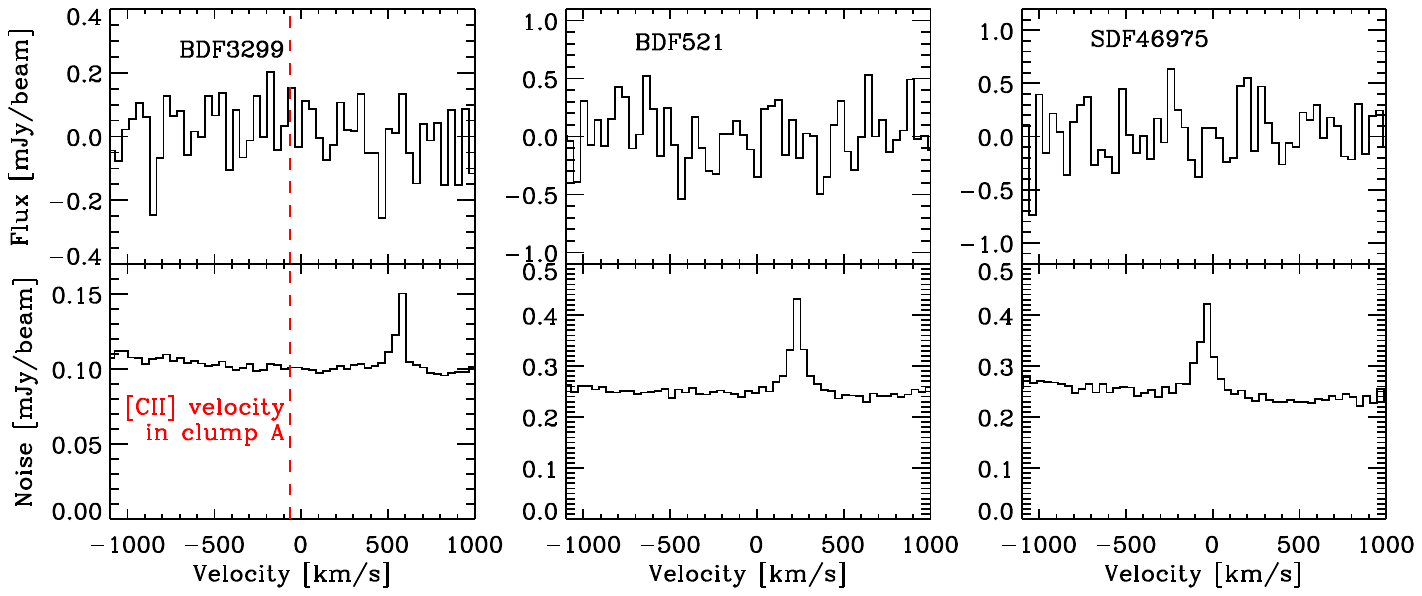}}
\caption{Top: spectra of the three sources in our sample, centered at the location of the optical
counterpart, extracted from the central beam. The velocity reference is set to the redshift defined by
the Ly$\alpha$ peak. Bottom: noise in the same velocity range measured across each plane associated with
each spectral bin. The regions where the noise is higher are associated with spectral regions of telluric
absorption. The spectral binning of these spectra is 40~km/s. The vertical red dashed line in the leftmost panel
indicates the velocity of the [CII] line detected next to BDF3299 in clump ``A'', illustrating that the line is located in a region of
low noise.}
\label{fig_spectra_nond}
\end{figure*}

\section{Sample selection, observations and data analysis}
\label{sect_obs}

The list of galaxies observed is given in Tab.~1.
These targets were initially selected as Lyman Break Galaxies through their z--Y dropout
\citep{Castellano2010,Ono2012} and all of them have solid spectroscopic
confirmations, through the detection of Ly$\alpha$ with clearly asymmetric profile due to
absorption of the blue side by the intervening IGM
\citep{Vanzella2011,Ono2012}.

The redshift given in Tab.1 refers to the redshift inferred from the peak of the
Ly$\alpha$ emission. This is probably significantly in excess (even by a few 100 km/s) with respect to  
the real systemic redshift of the sources, because of IGM absorption of the blue side of Ly$\alpha$,
and possibly also because of the effect of outflows \citep[e.g.][]{Pettini2002}.
The quoted star formation rates are based on their UV emission, assuming no dust extinction
\citep[which however is likely to be very low given the very steep UV continuum,][]{Vanzella2011}.
We have also corrected them, relative to the original papers,
by adopting the new relation provided by \citet{Kennicutt2012} (who use
a Kroupa IMF).

The ALMA observations were obtained between October 2013 and April 2014, with a number of antennae ranging
from 26 to 40, depending on the specific observation (by also taking into account the fact
that a few noisy antennae were removed --flagged as ``bad''--
from further reduction). An early observation of BDF3299
in May 2013 was not considered, because it was much noisier than the other data.
The antennae were distributed in a semi-compact configuration, with most of them within about
200m from the array center, but with one or two antennae located as far as 500-1000m.
The Precipitable Water Vapor during the observations ranged between 0.9 mm and 1.8mm,
with an average of about 1.2mm.

Observations were performed in Frequency Division Mode (FDM). Out of the four spectral windows, SPW1
was always centered on the expected frequency of the [CII] line, in the Upper Side Band. This spectral
band was set to a spectral resolution of 2.5~km/s. SPW0 was located on the continuum next to SPW1 (on
the higher frequency side), while SPW2 and SPW3 were located in the Lower Side Band to sample the
continuum. The spectral resolution of the ``continuum'' spectral bands
was different depending on the specific observation.

The phases were centered at the nominal optical position of the source (but see
discussion below about astrometric offset between
optical and millimeter data for BDF3299).

The band pass calibrators for BDF3299 and BDF512 were J2056-4714 and J2258-2758, while for SDF46975 the bandpass
calibrator was J1331+3030. The flux calibrator for BDF3299 and BDF521 was Neptune, while the flux calibrator for
SDF46975 were Ceres and Pallas. The phase calibrators for BDF3299 and BDF521 were J2223-3137 and J2247-3657,
while the phase calibrator
for SDF46975 was J1058+0133.

We note that the observations of BDF3299 were obtained in two different runs, one in October/November
2013 and one in April 2014. For the April 2014 observations, we moved the central frequency of SPW1 by about
100 km/s with respect to the observation in October 2013 and also the phase center was offset by about 0.6$''$.
Therefore, before combining the data, the data obtained in April 2014 were phase-shifted to match
the phase center of the observations in October 2013.

The ALMA data were reduced and analyzed by using CASA version 4.2.1.
Unfortunately this CASA version does not assign physically-based
weights to data taken at different epochs, with different integrations times, different atmospheric opacities
and with different spectral resolutions. As a consequence, before combining the visibilities
of two observations at 
two different epochs, we manually re-scaled their relative weights so that they matched the noise variance measured in the
two cubes. For the same reason, the same process had to be applied when combining data from different spectral
windows (e.g. to generate the continuum maps) taken with different spectral resolution, i.e. the weights of the different spectral
windows (when take with different resolutions) had to be adjusted manually.

The cubes were created with natural weighting and by shifting the channels to a common velocity reference
(i.e. by setting mode='velocity' in the CASA {\it clean} command).

The resulting cubes images typically have a beam size of about 0.7$''$, yet the detailed beam shape
and orientation for each observation is given in Tab.\ref{tab_list_sources}.
No cleaning was performed in any cube, since there are no sources bright enough in the field of view.
The only exception is the continuum image of BDF3299, in which a 400$\mu$Jy
serendipitous source is detected and which has enabled us to perform the cleaning (with 500 iterations)
of the continuum map.

The primary beam Full Width Half Maximum at this frequency is 22$''$.

The continuum maps were obtained by collapsing the four spectral windows (by taking care of the proper
weighting between spectral windows with different channel widths, as discussed above).

The final sensitivity reached in each set of data is given in Tab.1, both for the continuum
and for spectral bins (as detailed in the following).
In the deepest cube (the one on BDF3299) we reach a continuum sensitivity of 7.8$\mu$Jy by using the equivalent full
spectral band of 7.8 GHz.
The sensitivity to line detection depends on the specific frequency, since each band is affected by atmospheric
absorption in some spectral regions. The sensitivity also depends on the spatial location, both
because of the shape of the primary beam, which decreases the sensitivity at large radii from the phase
center, and because continuum serendipitous sources may generate sidelobes in their vicinity.
In Tab.1 we give the rms in spectral bins of 100~km/s obtained
by extracting the spectrum on the central beam (phase center)
and taking the rms spectrally, in the spectral region close to the expected frequency of the [CII] line.
The latter may require the selection of different spectral ranges depending of the source,
because of the different location of the
atmospheric absorption features in different spectra (relative to the expected [CII] frequency).
In the cases of BDF3299 and BDF521 the expected
location of the [CII] line is in an excellent atmospheric region. In these cases
we have measured the rms of the spectrum between --1000~km/s and +300~km/s; outside this range the rms increases
as a consequence of  atmospheric absorption features. In the case of SDF46975, an atmospheric
absorption feature is located just redward of the expected [CII] frequency from the Ly$\alpha$ redshift; since
we do not know exactly the expected location of the [CII] from the real accurate redshift, we
conservatively estimate the noise from the spectral range corresponding to the atmospheric absorption feature.

The absolute astrometric accuracy of the maps is given by the accuracy of the phase calibration
and the accuracy of the absolute coordinates of the phase calibrator. The global absolute astrometric
uncertainty is estimated to be about 0.15$''$. In the case of BDF3299, two observations were performed with two
different phase calibrators: in the two observations the continuum serendipitae have the same absolute position
within less than 0.1$''$, confirming the accuracy of the absolute astrometric calibration of ALMA.

The absolute astrometric
accuracy is only relevant for the comparison with the Y-band images. In the case
of BDF3299 the ALMA data reveal two serendipitous continuum sources detected at high level of significance \citep{Carniani2015}, both
of which have a clear counterpart in the Y-band image
or in the K-band image (which is cross-matched with the Y-band). However, both optical and near-IR counterparts are offset
by about the same amount in the same direction relative to the ALMA continuum sources, and
more specifically by $\sim 0.4''$ towards the NW
(PA$\sim 343^{\circ}$). Such a systemic offset is shown and discussed in Appendix \ref{appendix1}.
Since these offsets are the same for both continuum sources we believe they are associated with an offset
of the astrometry of the near-IR images relative to the ALMA images (we shall mention that
the absolute astrometry
of the Y-band and K-band images is linked to the USNO-B catalog, which has an average absolute astrometric accuracy of 
0.25$''$ at the epoch J2000, but variable across the sky with a dispersion about 0.12$''$).
Therefore, when comparing with the
Y-band image, we will apply to the Y-band image an offset of 0.4$''$ to the SE (PA$=163^{\circ}$).

\begin{figure}
\centerline{\includegraphics[width=8.5truecm,bb=40 157 541 641]{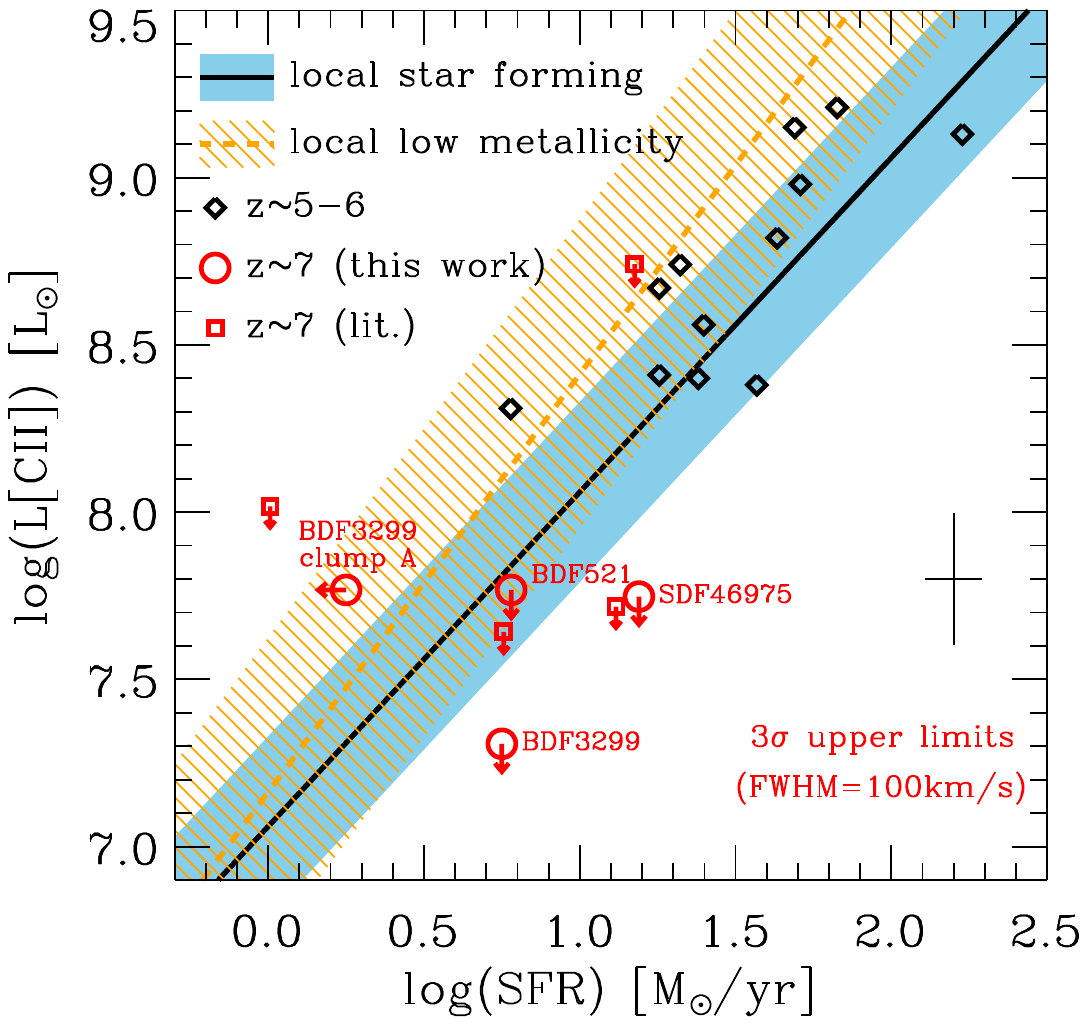}}
\caption{[CII]158$\mu$m luminosity versus SFR. The blue line shows the relation found by \citet{De-Looze2014}
for local star forming galaxies and starbursts (non including ULIRGs), while the shaded region shows the 1$\sigma$
dispersion. The orange dashed line and hatched region show the relation and dispersion found by \citet{De-Looze2014}
for local low metallicity dwarfs and irregulars.
The black diamonds are galaxies at $\rm z \sim 5-6$ from \citet{Capak2015} and \citet{Willott2015a} and their
average errorbar is shown in in the bottom-right of the diagram.
The red symbols are the [CII] data for galaxies at z$\sim$7
(specifically at 6.8$<$z$<$11): circles are our new observations and squares are data from the literature
\citep{Ota2014,Schaerer2015a,Gonzalez-Lopez2014}.
}
\label{fig_cii_sfr}
\end{figure}

\section{[CII] emission at the location of the optical sources}

The [CII] line is not detected (at a confidence higher than 3$\sigma$)
in any of the three sources at the location of their Y-band emission (i.e. within a radius
0.5$''$ from the latter).
We recall that in these sources the Y-band emission is partly contributed by UV stellar continuum emission
and partly by Ly$\alpha$ emission \citep[at least for BDF521 and BDF3299, Ly$\alpha$ and UV continuum contribute
to 1/3 and 2/3, respectively, of the observed Y-band flux,][]{Vanzella2011}.

The spectra extracted from the central beam are shown in Fig.~\ref{fig_spectra_nond},
along with the spectral distribution of the noise in the lower panels.

To estimate the upper limit on the line flux we assume a line width of $\rm 100~km/s$
\citep[we will see in the next section that this is the velocity width of the clump
detected next to BDF3299, moreover this width is consistent with other
lower redshift galaxies with low SFR, e.g. ][]{Williams2014,Capak2015}. Note that other authors
provide upper limits by assuming a line width of 40~km/s
\citep{Ota2014,Gonzalez-Lopez2014}; since these data will be used in our analysis, we have re-scaled their upper
limits to a line width of 100~km/s for consistency (this is simply done by
assuming that the noise on the line flux scales as the square root of the
line width).
The 3$\sigma$ upper limits on the [CII] luminosity are given in Tab.~\ref{tab_list_sources}.

Since [CII] is primarily excited as a consequence of heating by soft-UV radiation
(mostly in PDRs associated with star forming regions,
but also in HII regions or in the diffuse medium), many authors have identified a correlation
between SFR and [CII] luminosity \citep[e.g][]{Sargsyan2012,De-Looze2014}.

If the far-IR luminosity is used to measure the SFR, then the correlations seems to break down at high far-IR
luminosities (in the ULIRGs regime) and it is not yet clear whether this is associated with an
additional contribution to the far-IR from powerful AGNs, hence resulting in an inappropriate
measurement of the star formation rate from the far-IR emission \citep{Sargsyan2012,Malhotra2001},
or associated to different
physical conditions in ULIRGs, which may suppress the emission of [CII] \citep{Kaufman1999,Gracia-Carpio2011}.
However, our sources are certainly not in the ULIRG luminosity range, so we will not consider the
latter issue any further.

Fig.~\ref{fig_cii_sfr} shows the L([CII]) versus SFR relation for various classes of galaxies presented
in \cite{De-Looze2014}. The blue line indicates the relation for local normal star forming and starburst
(non-ULIRG) galaxies, with the shaded blue region giving the $\pm 1\sigma$ dispersion.
The orange dashed line and hatched region show the relation and dispersion of local low metallicity galaxies
 (in the range $\rm \frac{1}{40}<Z/Z_{\odot}<1$). For many of the low metallicity galaxies in the
 sample of \cite{De-Looze2014} the L([CII])/SFR ratio tends
 to be even higher than in normal galaxies (this is mostly a consequence of [CII] being a coolant of the ISM),
 however at very low metallicities ($\rm Z<0.1~Z_{\odot}$) the L([CII])/SFR start
 to decrease and becomes lower than in normal galaxies.

Galaxies at 5$<$z$<$6.1, from \citet{Capak2015}
and \citet{Willott2015a}, are shown with black diamonds and tend to follow the same [CII]/SFR relation as local galaxies.
These galaxies, with SFR typically of a few/several times 10~$\rm M_{\odot}~yr^{-1}$ are not yet representative of the bulk of the
population at these redshifts \citep{Salvaterra2011,Finkelstein2012,Dayal2014,Robertson2015}.

The 3$\sigma$ upper limits for ``normal'' galaxies at z$\sim$7 (those with
$\rm SFR < 100~M_{\odot}yr^{-1}$, i.e. not including the extreme QSO at z=7.08)
are indicated with red symbols in Fig.~\ref{fig_cii_sfr}. Circles indicate our new observations, while squares
are for additional galaxies at 6.8$<$z$<$11 from previously published studies
\citep{Ota2014,Schaerer2015a,Gonzalez-Lopez2014}.
At least three galaxies at z$\sim$7 for which a meaningful constraint on L([CII]) has
been obtained (including our ALMA deepest observation, BDF3299) have, at a given SFR, a 3$\sigma$ upper limit on their
[CII] luminosity significantly
lower than observed in most galaxies at lower redshifts, even by including the low metallicity ones.

However, we note that the lowest metallicity local galaxies in the \cite{De-Looze2014} sample ($\rm Z\sim
0.1-0.02~Z_{\odot}$) have [CII]/SFR values comparable with the $3\sigma$
upper limit on BDF3299 (although they have much lower SFR).
Therefore,
in principle, such a [CII] line deficit in z$\sim$7 normal galaxies could be ascribed to low metallicity.
Low metallicity of high-z galaxies has been a possible interpretation for the lack of [CII] detection in some star forming
galaxies at z$\sim$6.5 \citep{Ouchi2013,Vallini2013}.
Yet, one should bear in mind that the intensity of the [CII] line, as many other coolants
of the ISM, does not scale linearly with metallicity
\citep[see e.g.][]{Nagao2011,Madden2011,Cormier2012,rollig2006}.
Most importantly, we will show in the next section that
[CII] emission is actually
detected in a gas clump next to BDF3299. Therefore,
while metallicity may play a role, it is probably not
the only origin of the [CII] deficiency in the core of these primeval galaxies, as discussed in the following.

\begin{figure}
%{\includegraphics[width=8truecm,bb=0 220 540 720,clip]{cii_det_map_new.eps}}
%{\includegraphics[width=9.truecm,bb=69 204 554 842,clip]{cii_det_map_new.eps}}
{\includegraphics[width=9.truecm,bb=50 330 550 790,clip]{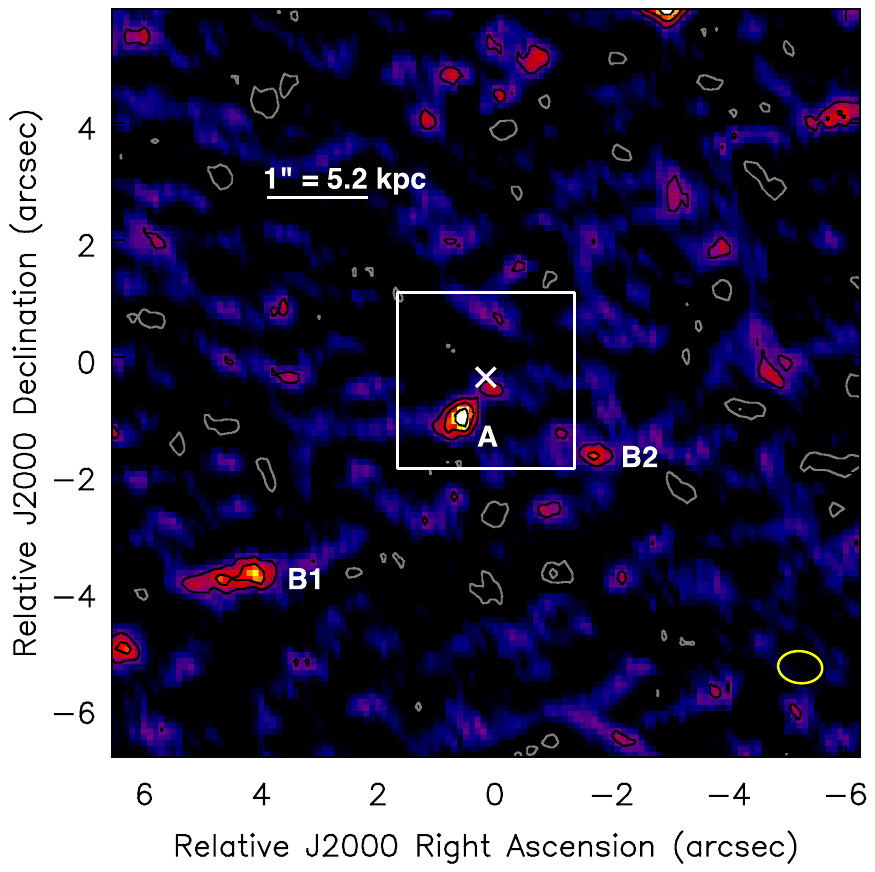}}
{\includegraphics[width=9.truecm,bb=50 330 550 790,clip]{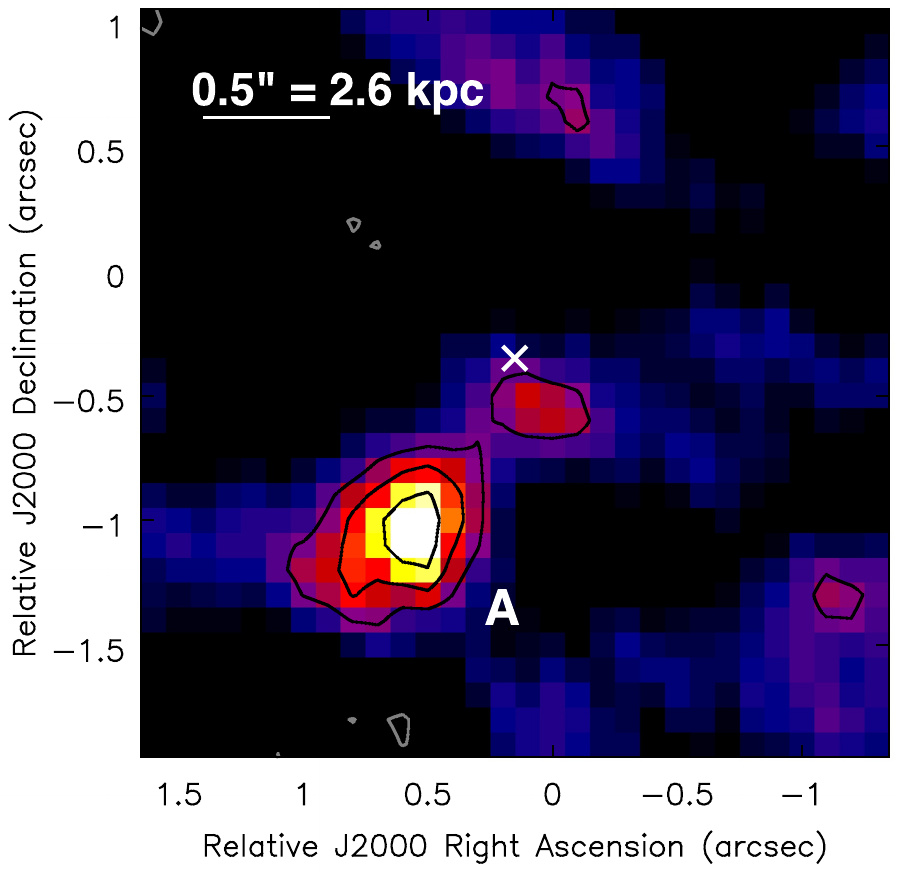}}
%\vspace{-2truecm}
\caption{Top: Map of the [CII] emission obtained with a velocity interval of 100~km/s, centered
at --64~km/s (relative to [CII] frequency expected from the Ly$\alpha$ redshift, z=7.109), in the ALMA data of BDF3299.
Black contours are at levels of 2, 3 and 4 times the noise per beam in the same map, i.e.
$\rm 6.4~~mJy~beam^{-1}~km~s^{-1}$. Grey contours are at the same levels, but for negative fluxes (note that $-4\sigma$ is
not reached anywhere). The ALMA
beam is indicated in the bottom-right corner. The white cross indicates the location of the Y-band image centroid
of the galaxy BDF3299. Label ``A'' indicates the contours associated with the [CII] detection, whose spectrum is shown in
Fig.~\ref{fig_cii_spec}, while ``B1'' and ``B2'' indicate two marginal detections discussed in the text.
Bottom: zoom of the central 3$''$ around the location of BDF3299. Note that the coordinates are relative to the phase center
of the ALMA observation in 2013, which was centered on the nominal coordinates of the optical image (but see the
text for a discussion
on the astrometry of the original Y-band images and their shift relative to the ALMA astrometry, which has been corrected here).}
\label{fig_cii_map}
\end{figure}

\begin{figure}
{\includegraphics[width=7truecm,bb=33 76 564 771]{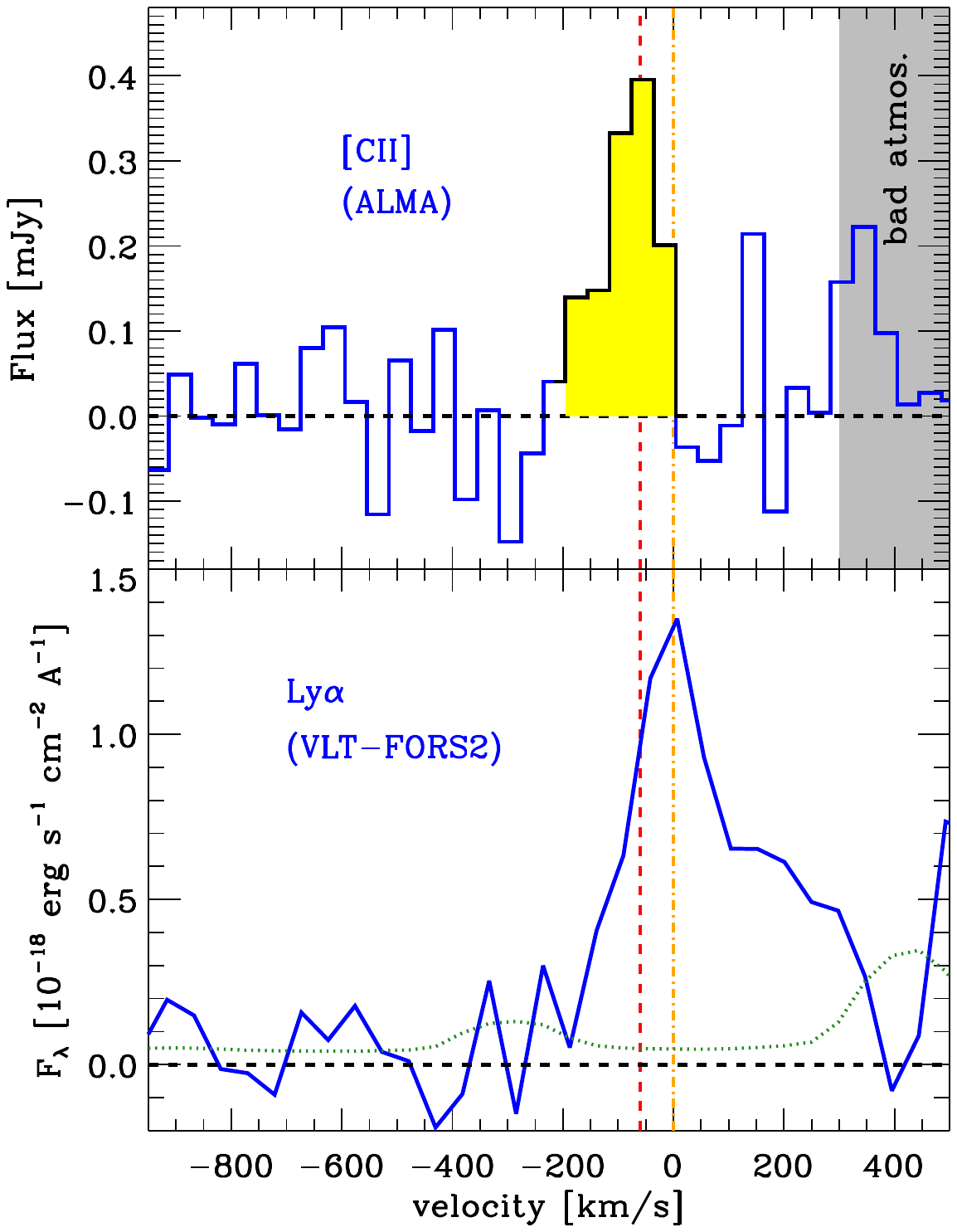}}
\caption{Top: ALMA spectrum of the [CII] source (clump ``A''), extracted from the an elliptical aperture obtained
by fitting the emission in the [CII] map (the spectrum flux has been corrected for the flux lost outside the extraction aperture, see text).
The gray shaded region indicates the part of the spectrum affected by
higher noise because of atmospheric absorption. Bottom: Ly$\alpha$ profile from the optical spectrum
obtained by \citet{Vanzella2011}. It should be noted the strongly asymmetric profile due to IGM absorption of
the blue side of the line. The green dotted line indicates the sky background spectrum. Velocities are
relative to the peak of Ly$\alpha$ (orange dot-dashed vertical line). The red dashed
line indicates the velocity of the peak of the [CII] line.}
\label{fig_cii_spec}
\end{figure}

\begin{figure}
%{\includegraphics[width=9.truecm,bb=69 205 554 842,clip]{overlay2.eps}}
{\includegraphics[width=9.truecm,bb=50 330 550 790,clip]{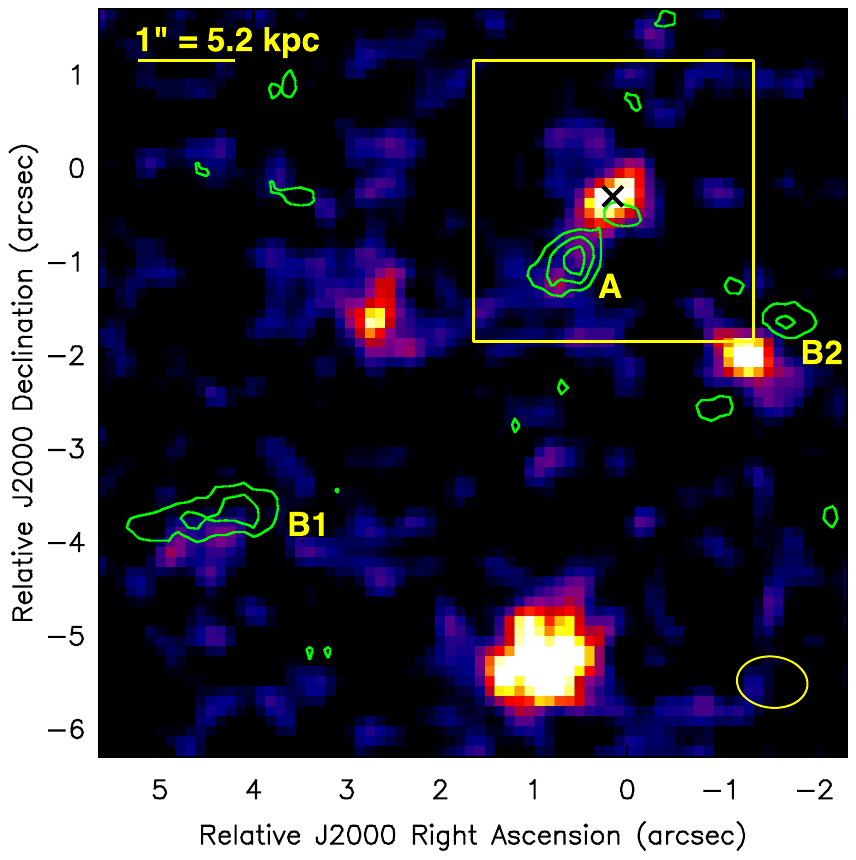}}
{\includegraphics[width=9.truecm,bb=50 330 550 790,clip]{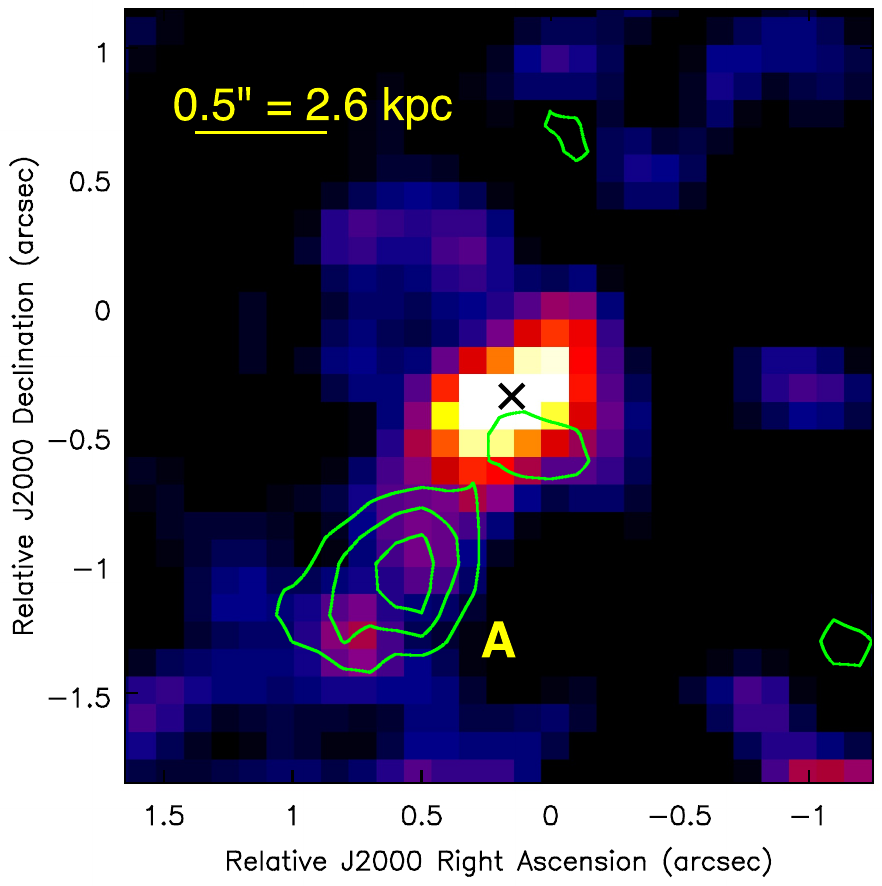}}
%\vspace{-2truecm}
\caption{Top panel: The color background image is the Y-band image of the BDF3299 field.
The black cross indicates the location of the galaxy BDF3299 at z=7.109. We recall that at this redshift
the Y-band samples both Ly$\alpha$ and UV continuum. The green contours show the [CII] map with the same
levels as (black contours) in Fig.~\ref{fig_cii_map}.
Label ``A'' indicates the contours associated with the [CII] detection, whose spectrum is shown in
Fig.~\ref{fig_cii_spec}, while ``B1'' and ``B2'' indicate two marginal detections discussed in the text.
The ALMA beam is shown in the bottom-right corner. Bottom panel: zoom of the central 3$''$ around the location of BDF3299.
Note that the coordinates are relative to the phase center
of the ALMA observation in 2013, which was centered on the nominal coordinates of the optical image (but see the
text for a discussion
on the astrometry of the original Y-band images and their shift relative to the ALMA astrometry, which has been corrected here).}
\label{fig_cii_y}
\end{figure}

\section{[CII] emission next to BDF3299 at z=7.1}

\subsection{Line detection}
\label{sec_line_det}

In the ALMA cube of BDF3299, which is the deepest of our observations,
we have detected a line consistent with [CII] at the redshift, z=7.1, inferred from the Ly$\alpha$ of the galaxy
(especially once IGM absorption of Ly$\alpha$ is taken into account), but
offset by 0.7$''$
(i.e. 4~kpc) relative to the 
location of the primary galaxy in the Y-band image (which traces UV+Ly$\alpha$ emission).

The map of the line emission
is shown in Fig.~\ref{fig_cii_map}, extracted with a spectral width of 100~km/s and centered at --64~km/s, relative to the
velocity scale set by the
redshift inferred from Ly$\alpha$.
The white cross indicates the centroid of the Y-band image of BDF3299.

The [CII] emitting clump is marked with an ``A'' and is located
at RA(J2000)=22:28:12.325, DEC(J2000)=--35:10:00.64. Its
emission peak is detected at 4.5$\sigma$.
However, the source is
marginally resolved (along a Position Angle of -33.5$^{\circ}$) hence the significance cannot be inferred
simply by the peak surface density in units of flux per beam. In this case, as for extended sources, to properly assess the significance of the
detection, we have to extract the spectrum of the source.
We have first fitted a two-dimensional Gaussian to the [CII] emission, and then
extracted the spectrum from an aperture corresponding to the section of the ellipsoid at half maximum of
its peak (i.e. an ellipses which has minor and
major axes of 0.55$''$ and 1.03$''$ respectively). Since the extraction aperture has a size close to the beam, we have applied a flux aperture
correction (inferred from the flux extracted with the same aperture 
from the continuum serendipitae), which results to be a factor of 2.5. The resulting spectrum is shown in
Fig.~\ref{fig_cii_spec}, rebinned to 40~km/s, together with the optical
spectrum of Ly$\alpha$. The line has a peak at a velocity of $\rm -71~km/s$ relative to the 
Ly$\alpha$ peak\footnote{This velocity is the average between the velocity estimated from a Gaussian fitting
($\rm -64~km/s$) and the first moment of the line ($\rm -78~km/s$).}. As already mentioned,
such a velocity offset is expected. Indeed the Ly$\alpha$ is redshifted relative to the systemic velocity
as a consequence of IGM absorption (as clearly shown by its strongly asymmetric profile), and possibly also due to the
effect of outflows \citep[e.g.][]{Pettini2002}, hence
the real redshift of BDF3299 is certainly slightly lower than inferred from Ly$\alpha$. Therefore, the blueshift by
$\rm -71~km/s$ is fully consistent with the rest-frame of the galaxy.
We have measured the significance of the line by measuring the noise in the spectrum between $\rm -1000$~km/s and
$\rm +300$~km/s; this is a region with uniform, low noise, while outside this spectral range atmospheric
absorption makes the noise higher (see Fig.\ref{fig_spectra_nond}). To avoid any potential
baseline residual instability, we
have also subtracted a continuum fitted on the same spectral window (which however is consistent with zero
within the noise). By integrating the line under the shaded region
in Fig.~\ref{fig_cii_spec} ($\rm -200<V<0~km/s$), we obtain a significance of the detection of 7$\sigma$.
A summary of the [CII] line properties of clump A is given in Tab.~\ref{tab_cii_det}.

We note that the same map in Fig.~\ref{fig_cii_map} reveals two potential
additional marginal detections (B1 and B2), which are the only ones in the map
with significance higher than 3$\sigma$ within a radius of 6$''$ and will be discussed later on.

The spatial location of the [CII] detection relative to the Y-band source is shown in Fig.~\ref{fig_cii_y}, in which
the green contours are the same as the (black) contours in Fig.~\ref{fig_cii_map}, while the colored background
image is the Y-band image.

We note that at the location of the optical source the ALMA [CII] map shows a
some signal at the $2\sigma$ level. However, this signal is very marginal
and it will not be discussed any further.

\subsection{Reliability of the detection}

\begin{figure}
{\includegraphics[width=8truecm,bb=58 187 537 585,clip]{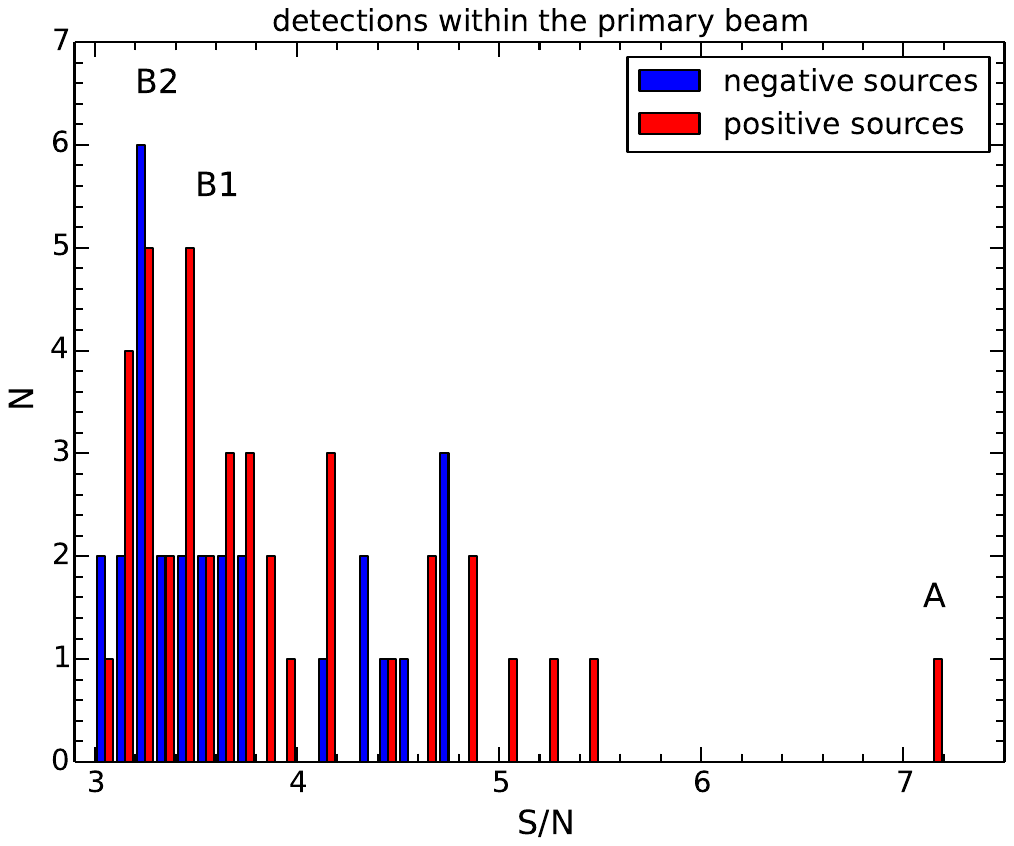}}\\
{\includegraphics[width=8truecm,bb=48 187 537 576,clip]{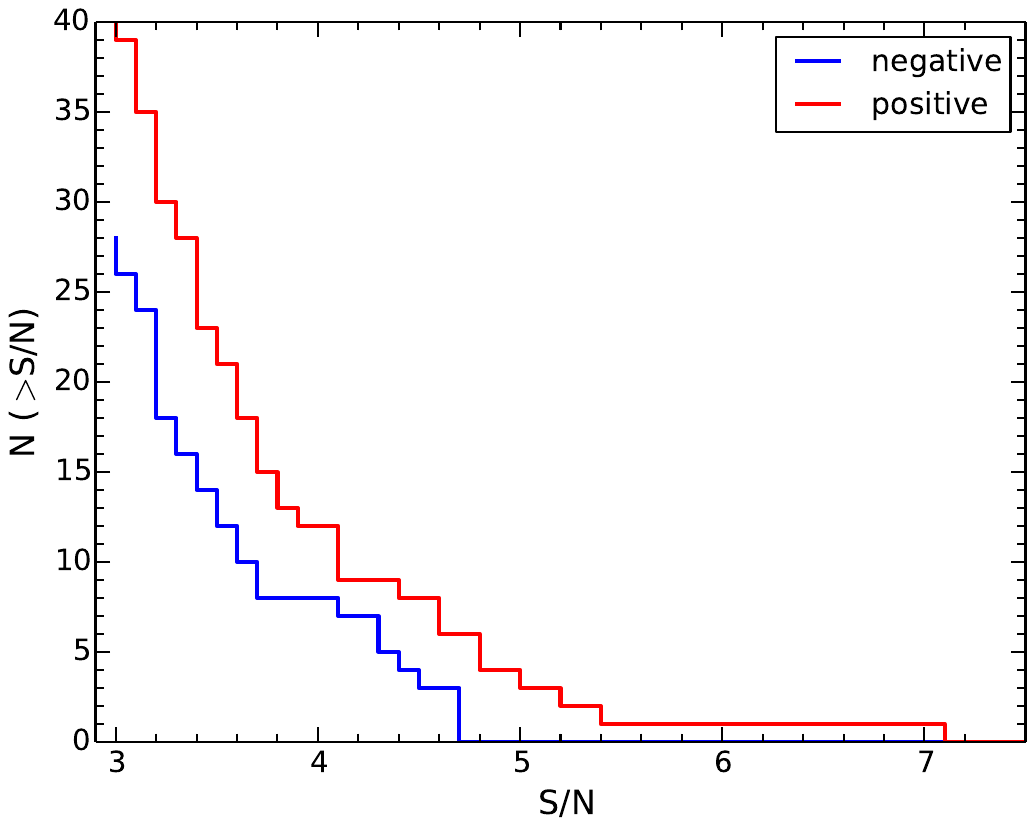}}\\
{\includegraphics[width=8truecm,bb=58 187 537 585,clip]{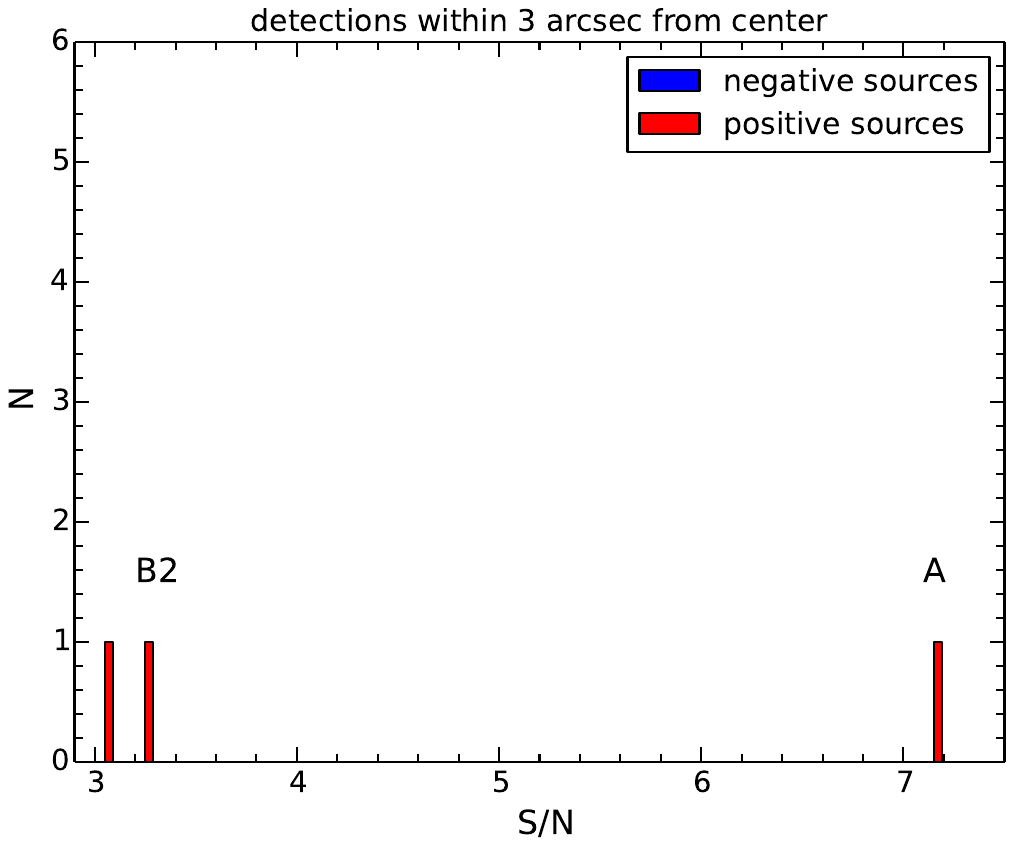}}
\caption{Top: Distribution of positive (red) and negative (blue) detections in the SPW1 of BDF3299, as a function of
signal-to-noise ratio of the detection, for lines integrated over the same velocity range as our [CII] detection,
searched wihtin one primary beam and in the velocity
range $\rm -1000~km/s<v<+300~km/s$ (i.e. within the range of good atmospheric transmission). Middle: cumulative distribution
of positive (red) and negative (blue) detections, within the same ranges. Bottom: Distribution of positive (red) and negative (blue) detections
within a radius of 3$''$ from the position of BDF3299, and over the same velocity range $\rm -1000~km/s<v<+300~km/s$.
}
\label{fig_pos_neg}
\end{figure}

In this section we discuss more in detail the significance of the [CII] detection.

First we have verified that the detection is
not the result of a glitch in a
single subset of the data. We have separated the visibilities obtained for BDF3299 in three data sets:
the set of observations obtained in November 2013, and we have divided into two halves the visibilities
obtained in April 2014. These three data subsets have similar depths (though the data in 2013 are somewhat
shallower than each of the two halves data taken in 2014). We have verified that in each of these three data
sets a signal of the [CII] line is seen, at a flux level consistent with the value observed in the merged data,
within the noise of each individual subset of data. More specifically, by only taking the (more sensitive) data
taken in 2014, the measured line flux is $\rm 46.0 \pm 7.7 ~mJy~km~s^{-1}$. If the 2014 data (4 scans in total)
are split in two groups of two scans each, then the line flux in each of the two scans is
$\rm 44.7 \pm 14.0 ~mJy~km~s^{-1}$ and $\rm 47.3 \pm 12.7 ~mJy~km~s^{-1}$. If only the (lower sensitivity) data
taken in 2013 are used, then the measured line flux is $\rm 49.9 \pm 17.9 ~mJy~km~s^{-1}$ (note that all of these
fluxes, and associated errors, have been scaled by the same aperture correction factor of 2.5 discussed above).

We have then checked whether negative sources are detected with the same significance or not. We have
searched for additional emission line detections within SPW1, within the spectral region between
-1000~km/s and +300~km/s (i.e. the region in which the noise is low and not affected by atmospheric
absorption features). We have adopted the same velocity width (100~km/s) as for the [CII] detection, to extract maps
for the initial selection of putative sources.
We have restricted the search to this criterion both
because the sidelobes of the continuum serendipitae may affect different velocity
ranges in a different way, and because we have verified that the noise does not scale exactly
as $\rm (\Delta V)^{-1/2}$, possibly because of some residual baseline problems (see also Vio et al., in prep, for a thorough
analysis of the noise in the ALMA data). We have discarded
putative detections (either positive or negative) for which:
1) the signal is not seen at the expected level (within the noise) in the three subsets of data discussed
above (i.e. the detection is mostly resulting from a glitch in one or two of the three subsets of data);
2) the measured size of the putative detection is smaller than
the synthesized beam, or 3) 
larger than 1.5 times the beam (real sources are not expected to be very extended). The latter two requirements
come from the fact that a source with size significantly smaller (or much larger) than the synthesized beam must
be due to noise associated with individual antennae, or group of antennae (these sources of noise are not correlated
and therefore should give features with angular profiles much different than the beam). Sidelobes of serendipitae, or
from strong sources outside the field of view, would also introduce features with sizes different from the beam.
However, to be on the safe side, we have also excluded the region within 2$''$ from the 0.4 mJy continuum serendipitous source, since the
uncleaned continuum map shows some low-level sidelobe residuals in this area.
For each putative detection passing this screening we have extracted the spectra as for the [CII] detection
and estimated the significance as discussed in Sect.\ref{sec_line_det} (i.e. over the same velocity interval of
200~km/s as for our detection). The resulting distribution of the positive (red) and
negative (blue) detections is shown, as a function of the Signal-to-Noise ratio,
in the top panel of Fig.~\ref{fig_pos_neg}. The cumulative distribution of the positive and negative detections (i.e.
the number of detection with significance above a given threshold) is shown
in the middle panel of Fig.~\ref{fig_pos_neg}. 

It is interesting to note that the number of positive detections at $\rm >3\sigma$
is 30\% higher than the number of negative detections, suggesting that a significant fraction of the positive
detections are real even down to the $3\sigma$ level. Focusing on the detections at $> 5\sigma$, we note that
there are no negative detections, hence giving further confidence that our positive detection at 7$\sigma$ 
next to the optical position of the galaxy (clump ``A''), is real.

If we restrict the search within a radius of 3$''$ from the optical location of
BDF3299 (but still ove the full velocity range
$\rm -1000 < \Delta V < + 300~km~s^{-1}$), then the resulting distribution of detections is shown in the bottom panel of  Fig.~\ref{fig_pos_neg}, in which there are
only three potential positive detections, and no negative detections, above 3$\sigma$. This further supports the
reliability of our detection in clump ``A''.

\subsection{Line identification}

The match of the line frequency with the one expected for [CII] at the redshift
of BDF3299 (Fig.~\ref{fig_cii_spec}) strongly supports this identification of the line. 

The identification with another transition associated with a foreground galaxy at lower redshift, which by
chance happens to be at 0.7$''$ from BDF3299, is extremely
unlikely, since the sky density of molecular and atomic lines in this band, and with fluxes comparable with
the line detected in our data, is extremely low. Even in such unlikely case, the galaxy should be clearly seen
in the optical/near-IR images and in the thermal millimeter continuum. In the following we quantify these
statements, hence excluding the line identification with a lower redshift transition.

The CO rotational transitions are the brightest transitions that can be seen in low/intermediate redshift
galaxies at this frequency. Other transitions are generally much fainter.
\cite{da-Cunha2013b} have inferred the cumulative space density of CO transitions as a function of flux
detection threshold.

We have considered their expectations including all CO rotational transitions from (7--6) to (2--1). \cite{da-Cunha2013b} consider both
the CO Spectral Line Energy Distribution (SLED) of the MW and the extreme CO SLED observed in the center of the starburst galaxy M82.
We have first conservatively adopted
the CO SLED of the M82 center; this is a conservative assumption, since the bulk of high-z galaxies have CO excitation
closer to the MW \citep[e.g.][]{Dannerbauer2009}.
With this assumption we obtain that the cumulative probability of finding a CO emitter within a radius of
1$''$ of a given position, within a velocity range of $\rm \pm 300~km~s^{-1}$ from a given frequency in band 6, and with a flux
brighter than 30~mJy~km/s (i.e. 50\% lower than the flux of the line detected by us), is about $\rm 10^{-4}$.

If the CO SLED typical of the MW (which is closer to the CO SLED of the bulk of high-z galaxies) is taken,
then the probability of chance CO detection drops to $\rm 2~10^{-5}$, within a velocity interval of $\rm \pm 300~km~s^{-1}$,
and to $\rm 7~10^{-6}$, within a more reasonable velocity interval of $\rm \pm 100~km~s^{-1}$.

In the calculations above we have conservatively assumed the putative foreground galaxies to have solar
metallicity, which has been used to estimate the CO sky densities. However, high redshift
galaxies are characterized by lower metallicity than local galaxies \citep[e.g.][]{Maiolino2008}. The CO-to-H$_2$ conversion
factor depends steeply on metallicity \citep{Bolatto2013a}, with low metallicity galaxies being much less efficient in
emitting CO for a given molecular gas mass. Therefore, the probability of finding a foreground CO emitter in the
vicinity of BDF3299, and at the same frequency expected for its [CII] emission, is even lower than inferred above.

In addition, to such extremely low probabilities of having a chance detection of a foreground, low-z CO emitter, one should
take into account that a putative CO emission, with the intensity observed by us, should be associated with a relatively
bright galaxy. Indeed, according to the Schmidt-Kennicutt law, CO emission tracing molecular gas in a galaxy, must
be associated to a certain SFR. Assuming, for instance, that the observed line is CO(4-3) at z=1 and conservatively assuming
the S-K relation observed for local galaxies \citep{Kennicutt2012}, one expects that the host galaxy should have
a $\rm SFR\approx 25 M_{\odot}~yr^{-1}$. Such a SFR should be associated with a galaxy with a brightness in the Y-band of
$\rm Y_{AB}=20~mag$. However, there is not such bright counterpart at the location of the line detection. Beside BDF3299 itself,
there is a faint source, marginally detected (Fig.\ref{fig_cii_y}), but confirmed in new HST images (Castellano et al. in prep.), located
0.3$''$ to the South-East, but which is {\it seven magnitudes fainter} than expected. The faintness in the Y band could in
principle be associated with extreme dust extinction. However, a SFR of $\rm 25 M_{\odot}~yr^{-1}$ should also result into a
10$\sigma$ continuum detection in our ALMA data, while we do not even have a marginal detection at the location of the line.
Similar or even more extreme expectations are obtained for other CO transitions or for any other submm molecular or atomic
transition.

These expected brightness in the Y-band and in the millimeter should be considered as very conservative. Indeed,
it has been shown that the S-K law evolves at high redshift, in the sense that for a given CO luminosity, or for a given
gas content, high-z galaxies are characterized by higher SFR than local galaxies \citep{Tacconi2013,Santini2014}.
High-z, low metallicity (faint) galaxies would be even more offset from the S-K relation,
in the sense that their CO emission would be even fainter (because of the steep metallicity dependence of the CO-to-H$_2$
conversion factor) or, equivalently, for a given CO luminosity the associated SFR rate (hence Y-band or far-IR continuum
luminosity) should be even higher than estimated above.

Finally, the arguments above, based on the S-K law, apply
to ``normal'' galaxies. Starburst galaxies have an efficiency of star formation (SFR per unit gas mass)
one order of magnitude higher \citep{Genzel2010a,Daddi2010b}. Therefore, depending on whether the putative foreground
galaxy is ``normal''
or ``starburst'' the expected luminosity in any band should be a factor of a few, up to an order of magnitude, higher than
estimated above.

Summarizing, the identification of the detected line with another transition associated with a foreground galaxy
is extremely remote, leaving [CII] at z=7.107 as the only realistic viable interpretation.

\begin{figure}
a)\\{\includegraphics[width=7truecm,bb=147 232 493 578]{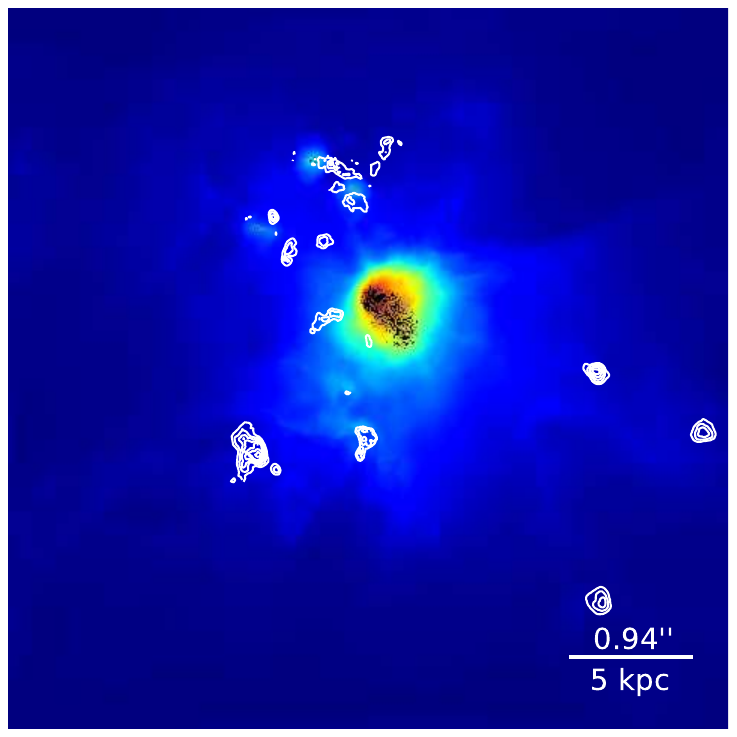}}\\
b)\\{\includegraphics[width=7truecm,bb=147 232 493 578]{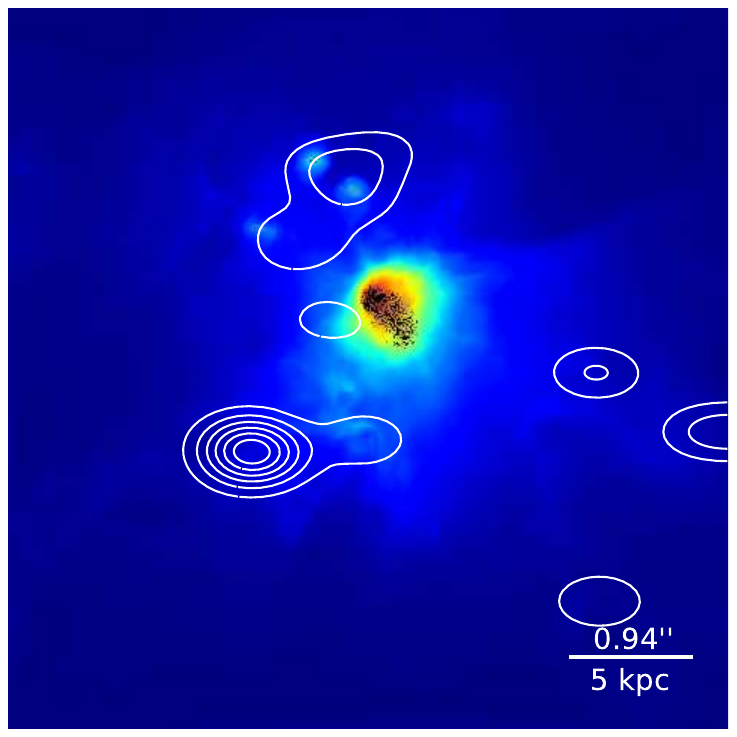}}\\
c)\\{\includegraphics[width=6.5truecm,bb=11 0 532 396,clip]{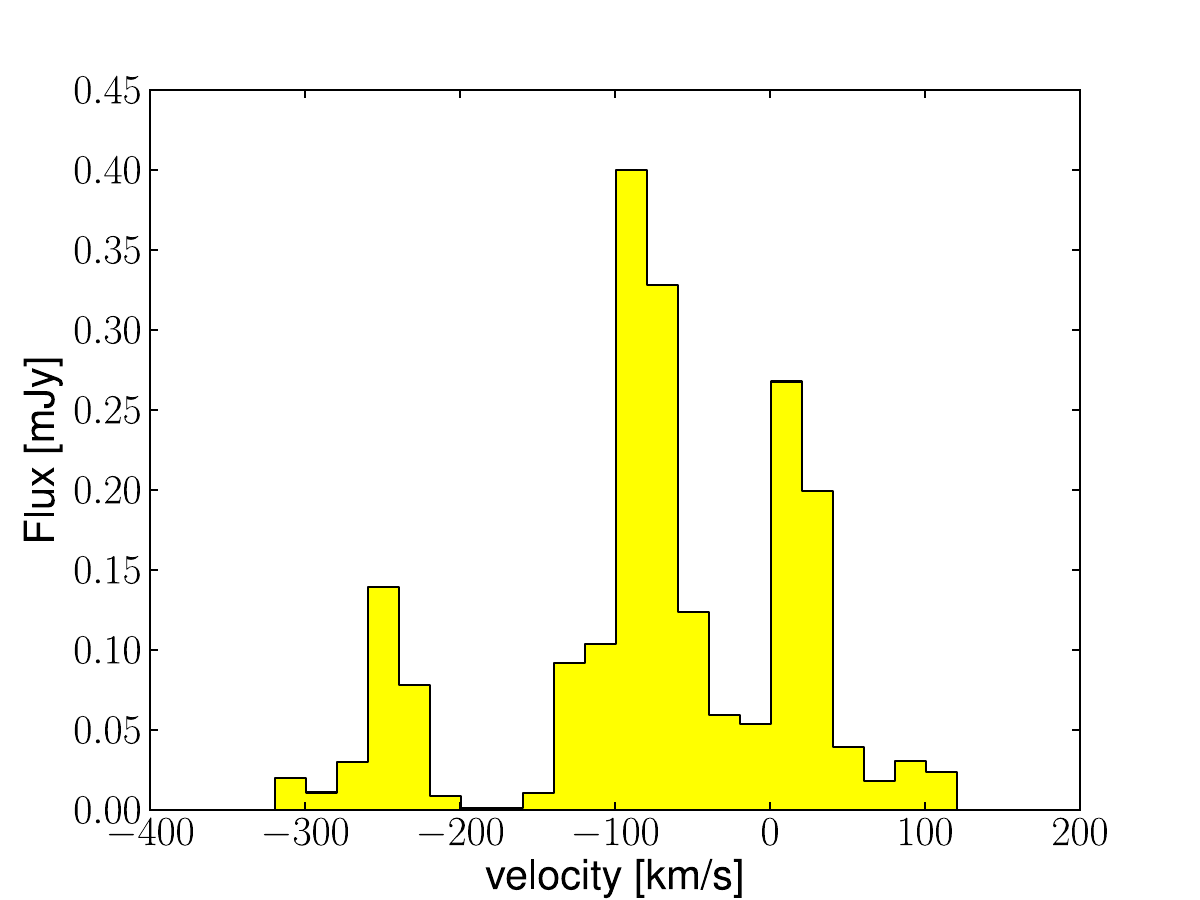}}
\caption{a) Simulation of a primeval galaxy at z=7.1
with SFR similar to BDF3299. The distribution of column density of ionized gas (which emits
Ly$\alpha$) is shown in colors (note that the observed diffuse Ly$\alpha$ emission can be more
extended than the warm gas as a consequence of scattering).
The black points show the distribution of young stars (UV continuum, which dominates the Y-band emission in our images). The
contours show the emission of [CII] tracing the neutral gas (distributed in clumps orbiting/accreting the primary
galaxy). b) Same as the top panel in which the image of the ionized gas has been convolved with the angular
resolution of our Y-band images (0.5$''$) and the [CII] map has been convolved with the ALMA beam of our data.
c) Integrated [CII] spectrum from the whole simulation.
}
\label{fig_vallini}
\end{figure}

\subsection{Interpretation of the [CII] emission}

[CII] emission offset relative to the UV and Ly$\alpha$ emission was already been seen in some other
high redshift galaxies \citep[e.g.][]{Williams2014,Gallerani2012}, therefore
our finding of offset [CII] emission is not a novelty. However, previous cases were associated with peculiar
objects, such as quasar host galaxies, or galaxies in an overdense environment and in the vicinity of a quasar.
More recently such displacement between [CII] emisison and UV/Ly$\alpha$ emission
has been observed also in high-z galaxies with more modest SFR
(a few/several times 10 $\rm M_{\odot}~yr^{-1}$), although still higher than the galaxies observed
by us, and at lower redshifts, 5$<$z$<$6.1 \citep{Capak2015,Willott2015a}.

Interestingly, displaced [CII] emission in primeval galaxies, as observed in our ALMA data,
is an expected consequence of the strong stellar feedback that, according to many models, should characterize galaxy
formation in the early Universe
\citep[e.g.][]{Salvadori2009,Dayal2014,Graziani2015}.
The [CII] line can be collisionally excited in many different ISM components  (e.g.
cold neutral medium, and high density photodissociation regions (PDRs) in molecular clouds). However, negative (thermal, radiative,
mechanical) feedback from stars can efficiently disperse or disrupt nearby molecular clouds, therefore quenching their [CII] PDR emission.
In this case, [CII] emission can only arise from accreting/satellite clumps of neutral gas displaced from the primary galaxy main body.
 
To illustrate this scenario more quantitatively, in Fig.~\ref{fig_vallini}a we show the model of primeval galaxies
developed by \citet{Vallini2013}, but specifically tailored to a galaxy at
z=7.1, with the same SFR as BDF3299.
In this model the gas in the main galaxy is completely ionized, while molecular clouds (and the associated PDRs)
are absent as a consequence of the strong stellar feedback
(details on the model are summarized
in Appendix \ref{appendix2}).
The black points show the distribution of young hot stars, i.e. what we observe as UV rest-frame continuum emission 
(we recall that the rest-frame UV-emission contributes to 2/3 of the observed Y-band emission).
The color image shows the distribution of warm ionized gas, which can be observed as
Ly$\alpha$ emission (note that the total observed Ly$\alpha$ emission can in principle be more
extended as a consequence of scattering).
The contours show the [CII]158$\mu$m
emission associated with the neutral gas (or mildly ionized gas). Clearly the central galaxy is
expected to emit strong UV continuum and Ly$\alpha$, but being
completely photoionized, it does not emit any [CII]. However, very interestingly, the model expects [CII] emission
from gas clumps at a few/several kpc ($\rm \sim 1''-2''$) from the primary galaxy. These are satellite gas clumps in the process
of accreting onto the primary galaxy, which have survived photoionization, owing to their distance from
the source of UV photons (and because a significant fraction of UV photons are absorbed internally by
the ISM of the primary galaxy).

Such [CII] emission in the satellite, accreting gas clumps is expected to be faint, but our deepest
observation, the one of BDF3299, does have the sensitivity to detect it.

In Fig.~\ref{fig_vallini}b we have smoothed the simulation to the angular resolution of our
optical observations (0.5$''$ seeing) and to the ALMA beam of our observations. The simulation
is not meant, by any means, to exactly reproduce our observations (this would require a huge number
of simulated objects with the same detailed modelling of the photoionization and radiative transfer in
the ISM and IGM shown here, and then finding the best matching case).
However, Fig.~\ref{fig_vallini}b illustrates that the offset between Y-band emission and [CII] emission
is indeed expected to be resolved for various [CII] clumps (and especially for the brightest one)
with the angular resolution delivered by our ALMA observation. Lower angular resolution observations
would probably hamper the capability of resolving the spatial offset and would probably associate the [CII] emission
with the primary UV-Ly$\alpha$ emitting galaxy.

The simulation also highlights that
more than one [CII] clump may be present around BDF3299, out to 3$''$ and possibly even at larger distances
(3$''$ is the maximum distance probed by the simulation box, so potential clumps at larger distances escape the simulation).
If present, these clumps are probably below our detection limit and may also suffer
beam dilution (many of the isolated small clumps disappear in the smoothed
simulated map).
Actually, in the ALMA cube of BDF3299 we find other positive [CII] detections, at lower significance than
clump ``A'', but in excess relative to
the ``negative'' detections, hence suggesting that they are not spurious detections.
Two of such marginal detections are marked as B1 and B2 in Fig.~\ref{fig_cii_map}. These are the only two
features with a peak flux detected at a significance higher than 3$\sigma$ within a radius of 6$''$ (half a beam
radius) from the primary galaxy, in the same map, centered at $\rm -64~km~s^{-1}$.
Certainly,
deeper and/or higher angular resolution observations are required to further confirm these marginal detections.

In principle the line width and spatial extension can provide information on the dynamical mass of the blob.
Determining a dynamical mass of the [CII] clump detected in the vicinity of BDF3299 is not easy, both because we do not know whether the line width
is tracing a rotating system or a velocity dispersion dominated system, and because the [CII] map is only
marginally resolved, hence the intrinsic dimension (beam deconvolved) is difficult to estimate.
If we assume that the system is rotationally supported and with a radius of about 1.5~kpc (inferred from the
Gaussian fit deconvolved from the beam in quadrature), we obtain a rough estimate of the clump dynamical
mass of about $\rm 5\times10^8~M_{\sun}$. If the system is not virialised, then this is actually an
upper limit to the dynamical mass.
Although all the dynamical mass estimate
is extremely uncertain, it suggests that indeed the [CII] clump is a small satellite of the
primary galaxy.

According to the model, in these satellite/accreting clumps
the [CII] excitation is not expected to primarily
originate from in-situ
star formation, but mostly as a consequence of heating from the soft-UV irradiation from the central primary galaxy.
This is consistent with the fact that we do not detect a clear Y-band source (i.e. Ly$\alpha$+UV continuum)
at the location of the [CII] clump. The vicinity of BDF3299 ($\rm Y_{AB}=26.15$) at 0.7$''$, and of another faint foreground galaxy
at 0.4$''$ ($\rm Y_{AB}\approx 27.5$, see next section), prevent
us to set tight upper limits on the Y-band emission associated with the [CII] clump, but it is certainly at least 3--4 times fainter
than the primary galaxy. According
to the [CII]-SFR relation for local (and intermediate redshift) galaxies \citep{De-Looze2014}, the [CII] luminosity of the clump
would imply a $\rm SFR \sim 6~M_{\odot}~yr^{-1}$, i.e. comparable to the main galaxy BDF3299, i.e. the [CII] clump
should have similar
or higher Y-band flux. The inconsistency between the [CII] clump and the local [CII]-SFR relation is also illustrated in
Fig.~\ref{fig_cii_sfr}, which also shows the location of the [CII] clump, undetected in Y-band, on the [CII]--SFR diagram.
The clump is clearly inconsistent with the distribution of local galaxies, but it
could still be marginally consistent with the expectations for low metallicity galaxies. Therefore, we cannot exclude some
contribution to the [CII] excitation from in-situ star formation, but we can also state that the non-detection of a clear
Y-band counterpart is certainly consistent with the models expectations.

An alternative interpretation could be that the rest-frame UV emission associated with putative star formation
in the [CII] clump is significantly obscured by dust. However, in this case one would expect associated dust emission
at the level of $\rm \sim 15 ~\mu Jy$, which is $\sim 2$ times higher than the rms of the continuum map, while we currently
do not even have a mariginal continuum detection.
We also note that the [CII] clump is not detected at longer wavelengths ($\rm J_{AB}>26.5$,
$\rm K_{AB}>26.0$). Significant dust content responsible for significant extinction has also been generally excluded by various
works on such high redshift
galaxies \citep{Walter2012,Ota2014,Schaerer2015a,Capak2015}, as also discussed in this paper (in Sect.~\ref{sec_cont}).
However, the current data are not deep enough to completely exclude the scenario of dust obscuration.
Additional data are certainly required to further dismiss or validate this possibility. Deeper
HST data are currently being obtained and future JWST data will certainly provide a definitive answer.

Obviously the clumps need some minimum level of metal enrichment to emit [CII].
The model suggests a metallicity
of about 0.05~Z$_{\odot}$. However, the metallicity is very difficult to constrain by using
a single emission line and especially based on [CII], since this one of the primary coolants of the
ISM. Therefore, the metallicity of the neutral gas clumps is highly uncertain and additional data are required
to properly constrain it.  Here we only note that a small level of metal enrichment in this clump may result either
from pre-enrichment as a consequence of metals expelled from the primary galaxy, or from in-situ production.
Indeed, even a very low SFR of only $\rm 0.2~M_{\odot}/yr$, certainly undetectable
by our observations (as discussed above), would be enough to enrich this small clump to a metallicity of $\sim$0.05~Z$_{\odot}$
in less than 100~Myr.

Fig.~\ref{fig_vallini}c shows the integrated [CII] spectrum from the simulation.
We note that the integrated [CII] spectrum inferred from the simulation, shown in Fig.~\ref{fig_vallini}c,
expects additional narrow emission [CII] lines emitted from small accreting/satellite
clumps illuminated by the primary source. We have tentatively identified
narrower (10-20~km/s) [CII] emission sources in the vicinity of BDF3299, but which
need to be confirmed with deeper and higher angular resolution observations.

The ALMA data of the other two sources in our sample (BDF521 and SDF46975) do not reveal clear evidence
of [CII] emission in the vicinity of the UV-Ly$\alpha$ primary galaxy. However, the latter data are
significantly shallower than the observation of BDF3299 (at least a factor of two in sensitivity).
The same line seen in BDF3299 would be undetectable in the lower S/N observations of BDF521 and SDF46975.

Moreover,
these two sources are brighter in the UV relative to BDF3299, hence their photoionization effect may extend
to larger distances, therefore preventing neutral clouds to survive in the proximity of the galaxy.

\subsection{Possible lensing from foreground galaxies}

We already mentioned that a faint galaxy, marginally detected in the Y-band (but
confirmed by recent HST data, Castellano et al. in prep.), is found at 0.3$''$ to the SE of the [CII] emitting
clump. The galaxy is marginally detected also in the V band, so it must be a foreground galaxy, certainly not
associated with the z=7.1 system. Proper constraints on the redshift of this foreground galaxy will come from
additional HST observations, as well as from forthcoming KMOS-VLT observations. Here we note that finding
a foreground galaxy so close to the ALMA line detection is intriguing.

The possibility that the line is actually CO at low redshift, and associated with this foreground galaxy, was already
discussed and discarded above: both the extremely low sky density of CO emitters and the extremely faint counterpart
(both in the Y-band and in the millimeter continuum) make this scenario implausible. Here we only add, to the
arguments discussed above, that the
putative CO emission would be offset from the Y-band counterpart, which would imply molecular gas offset by several
kpc from the galaxy center, which would be at odds with any known galaxy, where the molecular gas is found within the
central few kpc of galaxies.

It is possible that the proximity of this source is simply resulting from a chance superposition. The source has a magnitude
$AB\sim 27$ and density of sources at such faint fluxes is about $\rm 100~arcmin^{-2} mag^{-1}$ \citep{Guo2013}.
The probability of randomly finding a galaxy with this magnitude, or brighter, within a radius of 0.3$''$ of a
given position is less than $\rm 10^{-2}$. This probability is larger than that estimated for the chance CO
detection discussed above, but still relatively small.
%It is curious that the first ALMA detection of [CII] in
%the vicinity of a normal galaxy at z$>$7 is found in the vicinity of a foreground galaxy, which has a random chance
%probability of being found there of less than 1\%.

Another possible scenario is that the foreground low-z galaxy is gravitationally magnifying the [CII] emission
at z=7.1. Indeed, according to models, the primary galaxy at z=7.1 should be
surrounded by several [CII] emitting clumps. Although these clumps are abundant within a radius of $\sim 6''$ of the
primary galaxy (30~kpc), the [CII] emission is probably
below the detection threshold of our observations for most of them, especially as a consequence of their low
metallicity. If a foreground galaxy happens to be in the field, this can introduce an even mild gravitational
lensing that can boost the [CII] flux above our detection threshold.

This scenario would naturally explain the proximity of [CII] emission at z=7.1 with a foreground galaxy,
without invoking very unlikely chance overpositions. Indeed, in this scenario the several [CII] clumps become
visible (their emission is boosted above the detection threshold)
only when a foreground galaxy is located next to them in projection, so
that some gravitational lensing can occur.

The effect is particularly effective
if the foreground galaxy is
at z=1.5, which is the redshift maximizing the lensing magnification of radiation at z=7.1.
Unfortunately we do not have elements to test this scenario more quantitatively, since we do not yet have information
on the redshift, mass and light profile of the putative lensing galaxy and we do not have high enough resolution ALMA
observations to properly investigate the morphology of the putative lensed emission. Therefore, for the time being this
has to be regarded as a speculation, which would explain the observations, but to be tested with additional data.

However, in support of this lensing scenario, we note that the other two marginal [CII] detections in the vicinity of
BDF3299 (B1 and B2 in Fig.\ref{fig_cii_map}), are also found within 0.4$''$ of two foreground galaxies, as seen in Fig.\ref{fig_cii_y}.
The galaxy next to B1 is only marginally detected in our Y-band image, but it is confirmed in new HST data
(Castellano et al. in prep.). The galaxy next to B2 is clearly detected and the data currently available give
an approximate photometric redshift $\rm z_{phot}\sim 1$, which is consistent with the lensing scenario.

The cumulative probability of having, by random chance superposition, three Y-band detections within 0.4$''$ of
three given positions (our [CII] detection ``A'' and the other two marginal [CII] detections ``B1'' and ``B2'')
is $\rm 10^{-6}$. The lensing scenario would instead naturally explain the co-location of [CII] emission and
foreground galaxies, without having to invoke extremely low chance probabilities.

Certainly additional data are required to confirm and quantitatively test the lensing scenario.

\begin{figure}
{\includegraphics[width=7truecm,angle=90,bb=18 88 547 743]{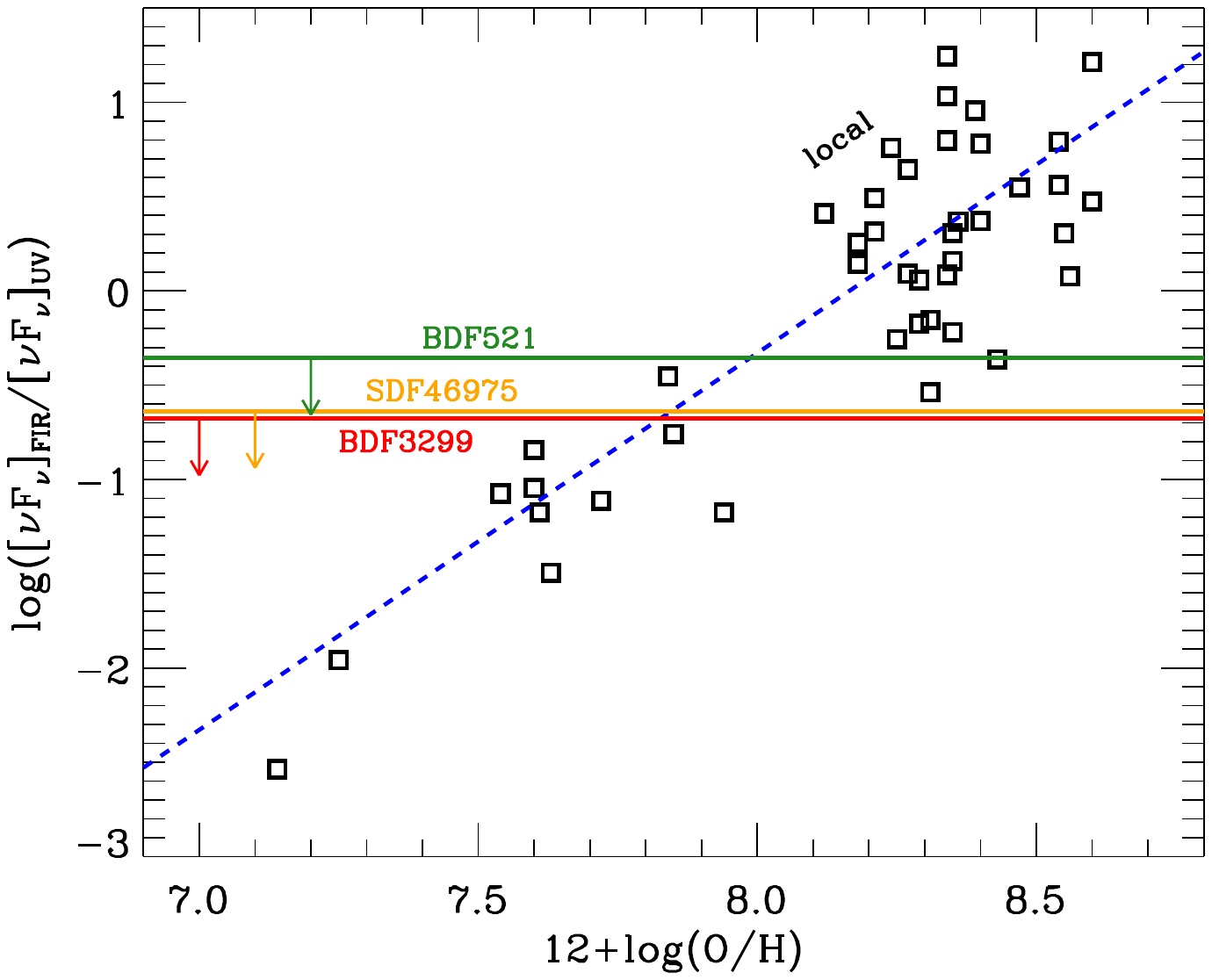}}
\caption{Far-IR ($\rm \lambda _{rest} = 160\mu m$) to far-UV ($\rm \lambda _{rest} = 1500$\AA)
luminosity ratio for local galaxies as a function of metallicity (black symbols). The blue dashed line is
a linear fit to the data. The colored solid lines indicate the upper limits on
$\rm (\nu F_{\nu})_{FIR}/(\nu F_{\nu})_{UV}$ inferred for the sample of our galaxies at z$\sim$7,
specifically: red for BDF3299, orange for SDF46975, green for BDF521.
}
\label{fig_uvir_oh}
\end{figure}

\section{Continuum dust emission}
\label{sec_cont}

While we have found a few serendipitous continuum detections \citep{Carniani2015},
the continuum dust thermal emission is not detected in any of our galaxies.
%The continuum images are shown in Fig.~\ref{fig_cont}.
This result is in line with previous results at high redshift
\citep{Walter2012,Ota2014,Schaerer2015a}, which
have found that
distant ``normal'' star forming galaxies, close to the re-ionization epoch,
have a rest-frame far-IR/UV luminosity ratio significantly lower relative to the bulk of local
star forming galaxies, with upper limits on the far-IR emission typically consistent with local low-metallicity
dwarf or irregular galaxies. This suggests that these early systems
are characterized by a small content of dust (which should
reprocess UV light into IR thermal emission),
both because of their low metallicity and because of shortage of time to produce dust
\citep{Valiante2009,Valiante2011,
Schneider2014,Hirashita2014,Nozawa2007}.
Our non detections go in the same direction, but achieving even tighter upper limits on the
far-IR/UV luminosity ratio.

Fig.~\ref{fig_uvir_oh} shows the ratio $\rm (\nu F_\nu)_{FIR}/(\nu F_\nu)_{UV}$ versus galaxy
metallicity for local galaxies, where $\rm (\nu F_\nu)_{UV}$ is estimated at
$\rm \lambda _{rest}=$1530\AA \ (i.e. GALEX FUV band, for galaxies in the local Universe), and
$\rm (\nu F_\nu)_{FIR}$ is estimated at $\rm \lambda _{rest}=$160$\mu$m (i.e. Herschel-PACS longest
wavelength band, for galaxies in the local Universe). Data are taken from \citet{Dale2007}, \citet{Dale2012},
\citet[][and priv. comm.]{Ota2014} and from \citet{Moustakas2010} \citep[in the latter we have used the calibrations
from ][]{Pilyugin2005}.
We have applied the correction to the far-IR fluxes, as discussed in \citet{Ota2014} and
\citet{da-Cunha2013a},
to take into account the fact that at high redshift
the detectability of thermal emission is more difficult because of the CMB temperature being
non-negligible relative to the dust temperature in galaxies (for the sake of simplicity we have
adopted the same correction assuming all three sources at z=7). For local galaxies,
there is a clear trend of $\rm (\nu F_\nu)_{FIR}/(\nu F_\nu)_{UV}$ to increase with metallicity,
implying that a larger fraction of the UV light is reprocessed by dust at high metallicities. This
is very likely a consequence of the dust content scaling with metallicity \citep[e.g.][]{Remy-Ruyer2014,
Draine2007a}.
The blue dashed line is a linear fit to the data in the log-log plane.
The 3$\sigma$ upper limits on the $\rm (\nu F_\nu)_{FIR}/(\nu F_\nu)_{UV}$ on z$\sim$7 sources are indicated
with horizonal colored lines (see legend).
Clearly, galaxies at z$\sim$7 are inconsistent with the UV-to-IR SED of typical local galaxies, with
solar-like metallicities, but are consistent with metal-poor (hence dust-poor) local dwarf galaxies,
with metallicities about 7 times lower than solar. Yet, while this suggests that the z$\sim$7 galaxies in our
sample must have metallicities similar or lower than local metal poor dwarfs, the latter typically have
SFR's one or two orders of magnitude lower than the sources observed by us. Therefore, the galaxies observed
by us at z=7.1 are probably scaled up versions (by at least an order of magnitude) of local low metallicity dwarfs.

Constraints on the dust mass are in principle
physically more interesting. However, it is not possible to obtain
an upper limit on the dust mass based on a single upper limit on the rest-frame far-IR or submm emission, since
the dust temperature is not known. \citet{Ota2014} assumes a dust temperature of 27.6~K (and
an emissivity index $\beta=1.5$) resulting from the average of 17 local dwarf and irregular galaxies.
\citet{Schaerer2015a} assume a range of temperatures between 25~K and 45~K. We follow a similar
approach as these papers to infer an upper limit on the dust mass from our 3$\sigma$ upper limits
at $\rm \lambda _{obs}=1.2~mm$, corrected for the effects of CMB.
Depending on the assumed dust temperatures we obtain upper limits on the dust mass in the
range between a few times $\rm 10^6~M_{\odot}$ and several times $\rm 10^7~M_{\odot}$,
as detailed in Tab.~\ref{tab_cont}. We note that
such low dust content is in agreement with the finding that the SFR inferred from Ly$\alpha$ is consistent, within
uncertainties, with the SFR inferred from the UV emission \citep{Vanzella2011}.

In the early universe the dust content is not only determined by the metallicity, but also
by timescale issues associated with the dust production mechanisms. Indeed, AGB stars, which are thought
to be the main factory of dust in the local universe
\citep[e.g.][]{ventura2012a,ventura2012b,di-criscienzo2013,Schneider2014},
require a timescale of several 100~Myr to evolve and to contribute significantly to the dust content,
hence a timescale comparable to the age of the universe at z$\sim$7. Dust forming in the ejecta of
core-collapse supernovae can be an additional source of dust on very short timescales, and indications
that SN dust can indeed contribute significantly to the overall dust budget in the early universe,
has been found
by observational studies \citep[e.g.][]{Maiolino2004b,gallerani2010,stratta2007,stratta2011}.
Extensive models have been developed to investigate the expected dust evolution in primordial galaxies
\citep[e.g.][]{Valiante2009,Valiante2011,Hirashita2014,Nozawa2015,Gall2011}.
Our tight upper limits on the dust mass in these early systems further suggest that at these early epochs
primeval galaxies have little time to produce masses of dust comparable with lower redshift galaxies.
A more quantitative comparison with models of dust formation in the early universe goes beyond the scope of
this paper and will be discussed in other forthcoming papers (Valiante et al. in prep., Ferrara et al. in prep.).

\begin{table*}
 \centering
 \begin{minipage}{140mm}
  \caption{Parameters of the [CII] detection next to BDF3299 in clump ``A''.}
  \begin{tabular}{cccccc}
  \hline
   $\rm \nu _{obs}([CII])^a$  &   z$_{\rm [CII]}$  & $\rm \Delta V^b$ & FWHM$^b$  &$\rm F([CII])$ &  L([CII]) \\
  			   $\rm [GHz]$	  &          & $\rm [km~s^{-1}]$ & $\rm [km~s^{-1}]$ & $\rm [mJy~km~s^{-1}]$ & $\rm [10^7~L_{\odot}]$ \\
 \hline
 234.43 & 7.107 & --71$\pm$10 & 102$\pm$21  & 48.6$\pm$6.9 & 5.9$\pm$0.8 \\
 \hline
\label{tab_cii_det}
\end{tabular}
\\
Notes:$^a$ Observed central frequency in LSRK. $^b$ Velocity offset relative to the peak of Ly$\alpha$ ($\rm \Delta V$)
and full
width half maximum (FWHM)
are calculated both through the moment analysis and through a Gaussian fit; the values reported in the table are average
values.
\end{minipage}
\end{table*}

\begin{table*}
\label{tab_cont}
 \centering
 \begin{minipage}{140mm}
  \caption{Constraints on the continuum upper limits.}
  \begin{tabular}{lccccccc}
  \hline
   Name     &   $\rm (\nu F_\nu)_{FIR}/(\nu F_\nu)_{UV}$$^a$  & $\rm M_{dust}(T=27.6K)$ &  $\rm M_{dust}(T=45K)$ \\
       &            & $\rm [10^7 M_{\odot}]$ &  $\rm [10^7 M_{\odot}]$ \\
 \hline
 BDF-3299 & $<$0.21 & $<$2.1 & $<$0.32 \\
 BDF-512  & $<$0.44 & $<$5.2 & $<$0.77 \\
 SDF-46975  & $<$0.23 & $<$5.8 & $<$0.87 \\
 \hline
\end{tabular}
\\
Notes:$^a$ $\rm (\nu F_\nu)_{FIR}$ is estimated at $\rm \lambda _{rest}=160\mu m$, $\rm (\nu F_\nu)_{UV}$ is estimated at $\rm \lambda _{rest}=1500$\AA. 
\end{minipage}
\end{table*}

\section{Conclusions}

We have presented new ALMA millimeter
observations aimed at constraining the [CII]158$\mu$m and thermal dust continuum emission
in three primeval galaxies at z$\sim$7, whose redshift had been
spectroscopically confirmed through the detection of (asymmetric) Ly$\alpha$.
In contrast to many previous millimeter observations of high redshift galaxies, which have targeted
galaxies with extreme star formation rates (100--1000~$\rm M_{\odot}~yr^{-1}$), the galaxies in our
sample have $\rm SFR \sim 5-15 M_{\odot}~yr^{-1}$, more typical of the bulk of galaxies at these early epochs.

At the location of the optical (Y-band) counterpart, which samples the rest-frame Ly$\alpha$ and UV emission,
no [CII]158$\mu$m line emission is detected. For some of the galaxies the 3$\sigma$ upper limit on the [CII] luminosity is significantly lower
than what is observed in local galaxies with similar star formation rates, even including many metal poor local galaxies
The upper limit on the [CII]/SFR ratio in these galaxies is also lower than observed in galaxies at $\rm z\sim 5-6$.

However, in the deepest of our ALMA observations, targeting the highest redshift galaxy of our sample (BDF3299 at z=7.109) we do detect
[CII] emission, slightly spatially offset, by 0.7$''$ (4~kpc), relative to the optical (Y-band) counterpart.

The line is detected at 7$\sigma$. The level of confidence is verified by checking the detection in subsets of the same data,
by checking the absence of any negative detection at confidence level higher than 5$\sigma$ in datacube over the whole ALMA primary
beam and over a broad frequency interval, as well as the absence of any negative detection at a confidence level higher than 3$\sigma$ within a radius of 3$''$ from the optical
source.

The detection of [CII] spatially offset from the optical galaxy, along with the absence of [CII] emission at the location of
the optical counterpart, is in nice agreement with the predictions of recent models of galaxy formation in the early universe.
According to these models, strong negative stellar feedback in these primeval systems quickly destroys or disperses molecular clouds,
hence quenching the emission of [CII] (the bulk of carbon is therefore in higher ionization states). However,
[CII] emission arises from accreting/satellite clumps of neutral gas displaced from the primary galaxy main body.

Therefore, our ALMA observations are directly probing such early phases of galaxy formation, in which both stellar feedback
and gas accretion are at work.

In the accreting clump [CII] emission is likely resulting primarily as a consequence of heating by the strong UV radiation from
the primary galaxy. However, some contribution to the [CII] excitation from in-situ star formation cannot be excluded.

At larger distances from the same galaxy ($\rm \sim 2''-4''$, i.e. $\rm \sim 10-20~kpc$) we obtain two additional marginal [CII]
detections at the redshift consistent with Ly$\alpha$. These marginal detections
need to be confirmed with additional data, but their presence would be in line with the expectations of the models discussed
above, according to which several satellites clumps of neutral gas should be emitting faint [CII] line in the surrounding of the primary
primeval galaxy.

It is interesting that both the [CII] detection and the two marginal detections are located in the vicinity (within $0.3''-0.5''$)
of foreground galaxies at lower redshifts. We show that the association of the millimeter lines with other transitions at lower
redshift  (in particular CO transitions)
is extremely remote and inconsistent with the properties of the counterparts (in terms of SFR). Chance
superposition coincidence is possible, but very unlikely. We suggest, as an alternative explanation, that gravitational lensing by foreground
galaxies may boost the flux of the [CII] clumps located in the vicinity of the primary z=7.1 galaxy,
hence bringing them above the detection
threshold. Testing this scenario more quantitatively requires a deeper knowledge of the properties of the foreground galaxies
(redshifts, masses, mass profile distribution, etc...)
and higher angular resolution data of the [CII] emission. However, if confirmed, this scenario would indicate that lensing from
foreground galaxies at z$\sim$1--2 could be an effective way to detect faint neutral clumps of accreting gas in the vicinity
of primeval galaxies around re-ionization.

Continuum thermal emission is not detected in any of the three galaxies of our sample. The 3$\sigma$
upper limit on the (rest-frame) IR-to-UV continuum
emission ratio is consistent with the value observed for the lowest metallicity local galaxies, in which the dust content
is very low. Constraints on dust masses are difficult to infer, because only one single photometric
upper limit per source is available for the IR SED,
hence no constraints on the dust temperature are available. By assuming the dust temperature typical
of local low metallicity galaxies, we obtain $3\sigma$ upper limits on the dust masses of a few times $\rm 10^7 M_{dust}$.
The latter confirms that these early systems have a very low dust content, both because of low metallicity and because of
shortage of time to produce dust at these early epochs through some of the standard dust production channels.

\section*{Acknowledgments}

We are grateful to the UK ARC node for assistance in the preparation of the ALMA observations, for providing the calibrated data.
We are grateful to the UK ARC node, to the central ESO ARC node and to Leonardo Testi for helpful
discussions on the data analysis and for their tips on the proper use of the CASA software. 
We also thank Marijn Franx and Padelis Papadopoulos for useful comments. We thank Ilse De Looze for providing the
data from her sample of galaxies. We are grateful to Kazuaki Ota, for providing the infrared and UV data on low metallicity
galaxies and for comments on the paper. This paper makes use of the following ALMA data:
ADS/JAO.ALMA\#2012.1.00719.S and  ADS/JAO.ALMA\#2012.A.00040.S; which can
be retrived from the alma data archive:
https://almascience.eso.org/alma-data/archive.
ALMA is a partnership of ESO (representing its member states), NSF (USA) and NINS (Japan), together with NRC (Canada) and NSC and ASIAA (Taiwan), in cooperation with the Republic of Chile. The Joint ALMA Observatory is operated by ESO, AUI/NRAO and NAOJ.

\setlength{\labelwidth}{0pt}
\bibliographystyle{mn2e}
\bibliography{bibl3}
%\begin{thebibliography}{99}

%\end{thebibliography}

%\bsp

%\label{lastpage}

\appendix

\section{Continuum serendipitae and registration with the optical images}
\label{appendix1}

\begin{figure}
%{\includegraphics[width=8truecm,bb=0 220 540 720,clip]{cii_det_map_new.eps}}
%{\includegraphics[width=9.truecm,bb=69 204 554 842,clip]{cii_det_map_new.eps}}
{\includegraphics[width=9.truecm,bb=50 330 550 790,clip]{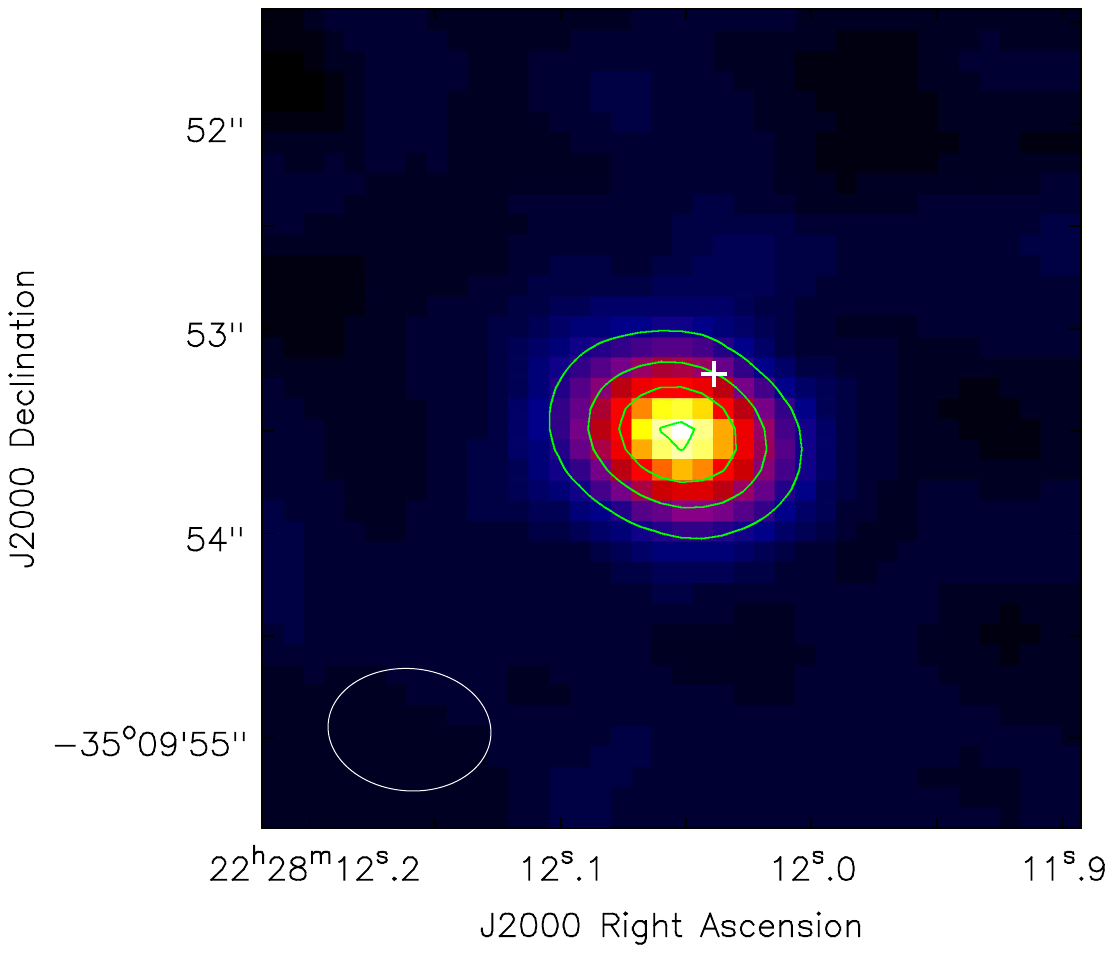}}
{\includegraphics[width=9.truecm,bb=50 330 550 790,clip]{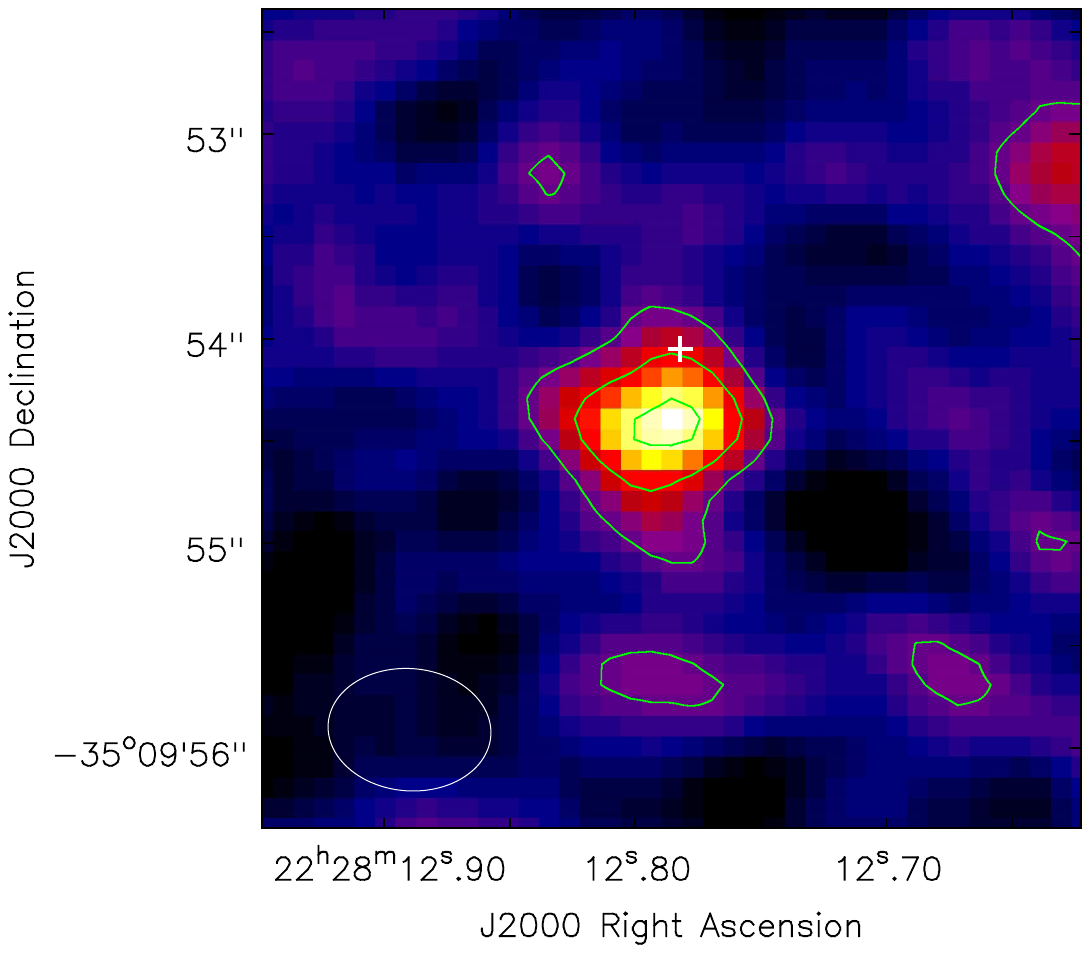}}
%\vspace{-2truecm}
\caption{Top: Continuum map of the brigest serendipitous source detected in the field of BDF3299. Contours are at
of 10$\sigma$, 20$\sigma$, 30$\sigma$, 40$\sigma$, where the rms=7.8$\mu$Jy. Bottom: Continuum map of the second
brightest serendipitous source in the same field. Contours are at
of 2$\sigma$, 4$\sigma$, 6$\sigma$. In both panels the white cross indicates the location of the near-IR counterpart.}
\label{fig_cont}
\end{figure}

The continuum map of BDF3299 reveals three serendipitous detections \citep{Carniani2015}:
two detections within the primary beam (one at $\sim 40 \sigma$ and another one at $\sim 6\sigma$)
and a third deetction outside the primary beam (at $\sim 4\sigma$).
Fig.~\ref{fig_cont} shows the continuum maps of the two detections within the primary beam.
Both detections have counterpart in the near-IR images (either Y-band or K-band). However, the centroid
of the near-IR counterparts is slightly offset relative to the millimeter emission in the ALMA map.
This is visible in  Fig.~\ref{fig_cont} where the white crosses show the location of the near-IR counterparts
(in images which are not yet registered as discussed in sect.\ref{sect_obs}).
Clearly both counterparts are offset by a similar amount and in the same direction
and more specifically by about 0.4$''$
towards the NW (PA$\sim 343^{\circ}$). Since these offsets are the same for both continuum sources
we believe they are associated with an offset of the astrometry
of the near-IR images relative to the ALMA images. We have therefore corrected for such an offset when comparing
the ALMA and near-IR images.

\section{The theoretical model}
\label{appendix2}

A detailed description of the model and simulations used in this paper is given
in \cite{Vallini2013}. In this appendix we only summarize some of the key points.

The \cite{Vallini2013} model is based on cosmological SPH hydrodynamic simulations performed with GADGET-2
\cite{Springel2005}. The original simulation reproduces a $\rm (10~h^{−1}Mpc)^3$ comoving volume with
$\rm 2 \times 512 ^3$ baryonic+dark matter particles, giving a mass resolution of
$\rm 1.32 \times 10^5~M_{\odot}$ for baryons and $\rm 6.68 \times 10^5~M_{\odot}$ for
dark matter. We select a snapshot at redshift z$\sim$7, and we
identify the most massive halo (total mass
$\rm M_h = 1.17 \times 10^{11}~M_{\odot}$, $\rm r_{vir} \approx 20~kpc$) by using a Friend-of-Friend algorithm. We then zoom the
simulation by selecting  a $\rm (0.65~h^{−1}Mpc)^3$ comoving volume around the center of the halo, and post-process UV radiative transfer
(RT) using LICORICE \citep{Baek2009}. Gas properties are resolved on a fixed grid with a resolution of 60 pc. To define the position of the ionizing sources we assume that stars form in those cells characterized by a gas density
$\rm \rho > 1~cm^{-3}$ in order to reproduce the typical size ($\sim$1--2 kpc) of star forming regions at z $\approx$ 6--7
\citep{Bouwens2004,Ouchi2009}, as inferred by UV continuum images. The population synthesis code STARBURST99
\citep{Leitherer1999} is used to obtain the ionizing spectrum of the galaxy.

The simulation is complemented with a sub-grid model taking into account the cooling and heating processes producing two-phase thermal
structure of the neutral gas in the ISM \citep[as in ][]{Wolfire1995,Wolfire2003}. The density ($\rm n_{CNM}$, $\rm n_{WNM}$) 
and temperature ($\rm T_{CNM}$, $\rm T_{WNM}$) of the the cold (CNM) and warm (WNM) neutral phase in each cell of the simulation is calculated as a function of :
1) the gas metallicity, Z, determining the coolants abundance, and
2) the Far- Ultraviolet (FUV) flux, G$_0$, in the Habing (6-13.6 eV) band, controlling the photoelectric heating produced by dust grains.
G$_0$ scales with the SFR and depends on the assumed age of the stellar population.
The density and temperature of the CNM and WNM allow us to compute [C II] emissivity from each cell of the simulated galaxy as in Eq. (2) in
\cite{Vallini2013}.

Variations of such model in which molecular gas and PDR are present in the main galaxy are presented in (Vallini et al. 2015,
submitted).

\end{document}